%% file: main.tex
\documentclass[prd,showpacs,letterpaper,twocolumn,nofootinbib,longbibliography,floatfix,superscriptaddress]{revtex4-2}

\input{Packages}

\input{Commands}

\begin{document}

\begin{CJK*}{UTF8}{gbsn}

%\preprint{IPPP/26/52}

\title{When CPT Violation Hides in Plain Sight: \\ How CP Measurements Are Compromised and How to Fix Them}% Force line breaks with \\

\author{Miaochen~Jin (靳淼辰)\orcidA{}}
\email{miaochenjin@g.harvard.edu}
\affiliation{Department of Physics \& Laboratory for Particle Physics and Cosmology, Harvard University, Cambridge, MA 02138, USA}

\author{Gabriela~Barenboim (加布里埃拉·巴伦博伊姆)\orcidC{}}
\email{gabriela.barenboim@uv.es}
\affiliation{Instituto de F\'{i}sica Corpuscular, CSIC-Universitat de Val\`{e}ncia, Paterna 46980, Spain}
\affiliation{Departament de F\'{i}sica Te\`{o}rica, Universitat de Val\`{e}ncia, Burjassot 46100, Spain}

\author{Carlos~A.~Argüelles\orcidB{}}
%\author{Carlos~A.~Argüelles (卡洛斯·阿尔圭列斯)\orcidB{}}
\email{carguelles@g.harvard.edu}
\affiliation{Department of Physics \& Laboratory for Particle Physics and Cosmology, Harvard University, Cambridge, MA 02138, USA}

\author{P.~Fern\'andez-Men\'endez(帕布洛·费尔南德斯)\orcidE{}}
\email{pablo.fer@cern.ch}
\affiliation{Universidade de Santiago de Compostela, Instituto Galego de Física de Altas Enerxías (IGFAE), Rúa de Xoaquín Díaz de Rábago, s/n, Santiago de
Compostela, 15705, Spain}

\author{I.~Martinez-Soler\orcidD{}}
\email{ivan.j.martinez-soler@durham.ac.uk}
\affiliation{Institute for Particle Physics Phenomenology, Department of Physics, Durham University, Durham DH1 3LE, UK}

\date{\today}% It is always \today, today,

\begin{abstract}
The extraction of the leptonic charge-parity (CP)-violating phase $\delta_{\rm CP}$
from long-baseline neutrino oscillation experiments rests on the assumption of charge-parity-time (CPT) conservation. We show that CPT violation, parametrized as an asymmetry $\delta\Delta m^2_{31} \equiv \Delta\bar{m}^2_{31} - \Delta m^2_{31}$ between neutrino and antineutrino mass splittings, induces an effective, energy-dependent phase shift $\phi_{\rm eff}(E)$ that is functionally degenerate with $\delta_{\rm CP}$ in the appearance asymmetry $\langle\Delta P\rangle$. This has a profound implication for long-baseline experiments, where the tension between T2K and NO$\nu$A CPT-conserving best-fit $\delta_{\rm CP}$ values can be significantly alleviated by a CPT-violating truth; and a CPT-conserving fit can miss the true CP phase entirely for $|\delta\Delta m^2_{31}|\gtrsim 0.3\times10^{-3}~\text{eV}^2$ for DUNE. We then demonstrate that atmospheric neutrino telescopes provide the natural tool to resolve this degeneracy: using existing data from IceCube-DeepCore (7.74 yr) and KM3NeT/ORCA-6 (433 kt-yr), we derive a world-leading constraint on CPT-violation at $| \delta\Delta m^2_{31}|\leq 0.57\times10^{-3}~\text{eV}^2$ at 90\% CL. With the IceCube Upgrade and full ORCA detector, we can reach a $1\sigma$ constraint at $10^{-4}~\text{eV}^2$ within a decade, providing the independent CPT constraint needed to ensure that DUNE's $\delta_{\rm CP}$ measurement is unambiguous.
\end{abstract}
\maketitle
% \tableofcontents
\end{CJK*}

% \tableofcontents

%%%%%%%%%%%%%%%%%%%%%%%%%%%%%%%%%%%%%%%%%%%%%%%%%%%%%%%%%%%%%%%%%%%%%%%%%

\section{Introduction}\label{sec:Intro}
\input{Sections/Intro.tex}

\section{Oscillations Phenomenology}\label{sec:pheno}
\input{Sections/Pheno.tex}

\section{Accelerators and Atmospheric Neutrino Telescopes}\label{sec:dune}

\input{Sections/Exp}

\section{Conclusion}\label{sec:conclusions}
\input{Sections/Conclusion} 

\section*{Code Availability and AI Usage}
The code used to perform analysis and to produce the plots in this paper are available at:

\textcolor{blue}{https://github.com/MiaochenJin/CPT-Hides-in-Plain-Sight}.

Large Language Models (LLMs) have been used in the development of this paper with heavy human supervision. All data analysis are end-to-end deterministic code pipeline (up to random seeds) and no text or code content has been created ground-up by LLMs. LLMs are used in structural and efficiency improvement of existing code architecture as well as slight language refining of human-written paper text and styling of existing plots. The code release is also rearranged from development-time code with LLM to maximize clarity and reproducibility.

\begin{acknowledgments}

 GB is supported by PID2023-151418NB-I00 funded by MCIU/AEI/10.13039/501100011033/, and by the European ITN project HIDDeN (H2020-MSCA-ITN-2019/860881-HIDDeN).
She also acknowledges support from the RCCHU.
PFM is supported by RYC2024-049537-I of the Spanish Ministry of Science.
CAA are supported by the Faculty of Arts and Sciences of Harvard University, Canadian Institute for Advanced Research (CIFAR), the National Science Foundation (NSF), the John Templeton Foundation, the Research Corporation for Science Advancement, and the David \& Lucile Packard Foundation.
MJ is funded by the National Science Foundation (NSF) CAREER Award \#2239795.
\end{acknowledgments}

%%%%%%%%%%%%%%%%%%%%%%%%%%%%%%%%%%%%%%%%%%%%%%%%%%%%%%%%%%%%%%%%%%%%%%%%%

% \nocite{*}
\newpage
%\clearpage
\bibliography{main}% Produces the bibliography via BibTeX.
\clearpage

%%%%%% SUPPLEMENTAL MATERIAL STARTS HERE
%%%%%% SUPPLEMENTAL MATERIAL STARTS HERE
%%%%%% SUPPLEMENTAL MATERIAL STARTS HERE
%%%%%% SUPPLEMENTAL MATERIAL STARTS HERE

\newpage

\onecolumngrid
\appendix

\ifx \standalonesupplemental\undefined
\setcounter{page}{1}
\setcounter{figure}{0}
\setcounter{table}{0}
\setcounter{equation}{0}
\fi

\renewcommand{\thepage}{Supplemental Methods and Tables -- S\arabic{page}}
\renewcommand{\figurename}{SUPPL. FIG.}
\renewcommand{\tablename}{SUPPL. TABLE}

\renewcommand{\theequation}{A\arabic{equation}}

\section{Details on Accelerator Experiments' Distributions and Measurements}\label{appx:appendix}

\input{Sections/Appendix-Acc}

\section{Details on Neutrino Telescopes' Distributions}\label{appx:appendix-ic}

\input{Sections/Appendix-IC}

\end{document}

%% file: Packages.tex
\usepackage{CJKutf8}
\usepackage{tikz}                    
\usetikzlibrary{arrows.meta, calc}   
\usetikzlibrary{angles, quotes}
\usepackage{adjustbox}
\usepackage{amsmath}
\usepackage{cancel}
\usepackage{graphicx}
\usepackage{fullpage}
\usepackage[colorlinks=true,allcolors=blue]{hyperref}
\usepackage[capitalise]{cleveref}
\usepackage[utf8]{inputenc}
\usepackage{siunitx}
\usepackage{array}
\input{units}

\usepackage{pifont}
\usepackage{enumitem}
\usepackage{url}
\usepackage{rotating}
\usepackage{epsfig,graphics,rotate}
\usepackage{flushend}
\usepackage{bm}
\usepackage{amssymb}
\usepackage{amsmath}
\usepackage{amsfonts}
\usepackage{gensymb}
\usepackage{booktabs}

\usepackage{microtype}
\usepackage{multirow, makecell}
\usepackage{lipsum}
\usepackage{fullpage}
\usepackage[normalem]{ulem}

\usepackage{xspace}
\usepackage{comment}
\usepackage{floatrow}
\usepackage{ragged2e}

\usepackage{nicefrac, xfrac}
\usepackage{mathtools} % for 'pmatrix' environment
\usepackage{relsize}   % for '\larger' macro

\usepackage{float}

%% file: units.tex
\DeclareSIUnit \s {\second}
\DeclareSIUnit \ns {\nano\second}
\DeclareSIUnit \mus {\micro\second}
\DeclareSIUnit \ms {\milli\second}
\DeclareSIUnit \MB {\mega\byte}
\DeclareSIUnit \GB {\giga\byte}
\DeclareSIUnit \TB {\tera\byte}
\DeclareSIUnit \PB {\peta\byte}
\DeclareSIUnit \Mbps {\mega\bit/\s}
\DeclareSIUnit \Gbps {\giga\bit/\s}
\DeclareSIUnit \Tbps {\tera\bit/\s}
\DeclareSIUnit \Pbps {\peta\bit/\s}
\DeclareSIUnit \kton {\kilo\tonne} % changed  back to kton
\DeclareSIUnit \kt {\kilo\tonne}
\DeclareSIUnit \Mt {\mega\tonne}
\DeclareSIUnit \eV {\electronvolt}
\DeclareSIUnit \keV {\kilo\electronvolt}
\DeclareSIUnit \MeV {\mega\electronvolt}
\DeclareSIUnit \GeV {\giga\electronvolt}
\DeclareSIUnit \TeV {\tera\electronvolt}
\DeclareSIUnit \PeV {\peta\electronvolt}
\DeclareSIUnit \EeV {\exa\electronvolt}
\DeclareSIUnit \m {\meter}
\DeclareSIUnit \cm {\centi\meter}
\DeclareSIUnit \in {\inchcommand}
\DeclareSIUnit \km {\kilo\meter}
\DeclareSIUnit \kV {\kilo\volt}
\DeclareSIUnit \kW {\kilo\watt}
\DeclareSIUnit \MW {\mega\watt}
\DeclareSIUnit \MHz {\mega\hertz}
\DeclareSIUnit \mrad {\milli\radian}
\DeclareSIUnit \year {years}
\DeclareSIUnit \POT {POT}
\DeclareSIUnit \sig {$\sigma$}
\DeclareSIUnit\parsec{pc}
\DeclareSIUnit\lightyear{ly}
\DeclareSIUnit\foot{ft}
\DeclareSIUnit\ft{ft}
\DeclareSIUnit \ppb{ppb}
\DeclareSIUnit \ppt{ppt}
\DeclareSIUnit \samples{S}
\DeclareSIUnit \pe{PE}

\newcommand{\enu}{\E_\enu}

%% file: Commands.tex
\definecolor{lime}{HTML}{A6CE39}
\DeclareRobustCommand{\orcidicon}{
	\begin{tikzpicture}
	\draw[lime, fill=lime] (0,0) 
	circle [radius=0.16] 
	node[white] {{\fontfamily{qag}\selectfont \tiny ID}};
	\draw[white, fill=white] (-0.0665,0.095) 
	circle [radius=0.005];
	\end{tikzpicture}
	\hspace{-2mm}
}

\foreach \x in {A, ..., Z}{\expandafter\xdef\csname orcid\x\endcsname{\noexpand\href{https://orcid.org/\csname orcidauthor\x\endcsname}
			{\noexpand\orcidicon}}
   }
   
\newcommand{\abar}{\bar \alpha}
\newcommand{\dCP}{\delta_{\rm{CP}}}
\newcommand{\phieff}{\phi_{\rm{eff}}}

%% file: Sections/Intro.tex
The discovery of neutrino oscillations~\cite{Fukuda:1998mi,Ahmad:2002jz} stands as one of the landmark results of modern particle physics, establishing that neutrinos are massive and that lepton flavors mix.
Yet the minimal extension of the Standard Model required to accommodate this fact leaves many fundamental questions unanswered: the absolute neutrino mass scale; the Dirac, Pseudo-Dirac, or Majorana nature of the mass; the mass ordering; and, most relevant to the present work, whether the leptonic sector violates CP symmetry.
The CP-violating phase $\delta_{\rm CP}$ enters the Pontecorvo--Maki--Nakagawa--Sakata (PMNS) mixing matrix and governs a matter--antimatter asymmetry in neutrino oscillations that next-generation long-baseline experiments, chief among them DUNE~\cite{DUNE:2020jqi} and Hyper-K~\cite{Abe2018Hyper-KamiokandeReport}, are specifically designed to measure.

A reliable extraction of $\delta_{\rm CP}$, however, rests on a tacit assumption that is rarely
scrutinized in practice: \emph{CPT conservation}. 
Long-baseline oscillation experiments are sensitive to the difference between neutrino and antineutrino oscillation probabilities, $\Delta P \equiv P(\nu_\mu \to \nu_e) - P(\bar\nu_\mu \to \bar\nu_e)$, which is sourced both by genuine CP violation and by any CPT-violating asymmetry between the mass spectra and mixing parameters of neutrinos and antineutrinos.
If CPT is violated at even a modest level, a CPT-conserving analysis will absorb the extra asymmetry into a biased inference of $\delta_{\rm CP}$, potentially reporting a spurious signal of CP violation or obscuring a real one.
This concern is not merely theoretical: it was shown in~\cite{Barenboim:2020tzf} that the hints of CP violation reported by T2K, a predecessor of Hyper-K, are not robust against a CPT-violating explanation, and that a CPT-violating, CP-conserving scenario is in perfect agreement with the neutrino oscillation data available at the time, even after combining T2K with NO$\nu$A and reactor experiments. 
Testing CPT conservation is therefore not merely an exotic pursuit; it is a necessary prerequisite for the program of discovery of leptonic CP violation.

Beyond its immediate practical consequences for oscillation physics, CPT symmetry occupies a singular position in the architecture of modern theoretical physics. 
CP violation, however striking, is accommodated within the Standard Model through the Cabibbo--Kobayashi--Maskawa and PMNS phases; its discovery refined the theory but did not overturn its foundations. 
CPT violation, by contrast, would constitute a far more radical departure. 
The CPT theorem~\cite{Luders:1954zz,Pauli:1955,Bell:1955,Jost:1957zz} guarantees that any theory satisfying three pillars of quantum field theory---Lorentz invariance, locality, and the spin-statistics connection in a unitary, local, relativistic framework---must be invariant under the combined operation of charge conjugation C, parity inversion P, and time reversal T.
The theorem does not require that any of C, P, or T be individually conserved; it demands only their product.
A confirmed violation of CPT would therefore signal the breakdown of at least one of these foundational principles: it would imply either a departure from Lorentz invariance, a non-local interaction structure, or a failure of the standard spin-statistics relation. 
Any of these would necessitate physics beyond the framework of local relativistic quantum
field theory---the mathematical language underlying every confirmed fundamental theory we possess.

Proposals for CPT violation arise naturally in several beyond-Standard-Model contexts. In string-field
theory and non-commutative geometry, spontaneous breaking of Lorentz symmetry generates CPT-odd
operators in the low-energy effective Lagrangian~\cite{Kostelecky:1988zi,Kostelecky:1991ak}. The
Standard-Model Extension (SME) of Kosteleck\'y and collaborators provides a systematic
effective-field-theory catalog of all Lorentz- and CPT-violating operators consistent with the
remaining symmetries of the Standard
Model~\cite{Colladay:1996iz,Colladay:1998fh,Kostelecky:2003fs}. A qualitatively different and
particularly minimal realization was proposed in~\cite{Barenboim:2001ac}: CPT violation in the Dirac
mass terms of the three neutrino flavors, which generates independent mass spectra and mixing
parameters for neutrinos and antineutrinos while \emph{preserving Lorentz invariance}. This scenario,
strongly motivated by braneworld models with extra dimensions where neutrinos propagate into the bulk,
parametrizes the CPT-violating observable most directly as $\delta\Delta m^2_{31} \equiv
\Delta\bar{m}^2_{31} - \Delta m^2_{31} \neq 0$, and is the framework adopted throughout this work.
Quantum-gravity scenarios involving space-time foam or decoherence induced by virtual black holes also
generically predict CPT non-invariance~\cite{Amelino-Camelia:1997ieq,Mavromatos:2004sz}. The neutrino
sector thus provides a low-energy window onto Planck-scale physics.

Neutrinos are, in several respects, the most powerful laboratory for CPT tests among elementary particles. 
First, unlike the kaon or $B$-meson systems, where CPT tests rely on particle--antiparticle
mass differences, oscillation experiments are directly sensitive to the interference between mass
eigenstates, giving access to the full matrix of mass-squared differences and mixing angles for both
neutrinos and antineutrinos separately. 
Second, the origin of neutrino mass remains unknown. If neutrinos are Majorana particles---their own antiparticles---the usual notion of a CPT-conjugate state requires re-examination; the relevant CPT-odd quantity becomes a difference in the effective Majorana masses or phases.
If they are Dirac, neutrino and antineutrino mass matrices are in principle independent, and an asymmetry $\delta\Delta m^2_{31}$ is a direct, gauge-invariant observable. 
In either case, the lack of a confirmed mass-generation mechanism means that new physics in the neutrino mass sector is not only possible but perhaps expected, and CPT-violating mass terms are among the most minimal such extensions. 
Third, cosmological and astrophysical neutrinos traverse macroscopic and even Hubble-scale distances, providing extraordinary sensitivity to Lorentz- and CPT-violating effects that accumulate coherently with
baseline~\cite{Kostelecky:2003cr,IceCube:2010fyu}.
Fourth, the atmospheric neutrino flux delivers a broad-band, isotropic sample of both neutrinos and antineutrinos spanning energies from sub-GeV to hundreds of TeV, and experiments such as IceCube-DeepCore and KM3NeT/ORCA are able to exploit this diversity of $L/E$ to disentangle CPT-violating mass asymmetries from matter effects and standard-oscillation parameters.
In fact, the next-generation atmospheric neutrino experiments will perform some of the most precise measurements of the neutrino oscillation parameters~\cite{Arguelles2023MeasuringNeutrinos}, including the CP-violating phase~\cite{Razzaque:2014vba,Kelly:2019itm,Denton:2023qmd,Beacom:2026mys}.

In this work, we study the interplay between CPT violation and CP violation in long-baseline neutrino oscillation experiments.
We demonstrate, analytically and numerically, that CPT-violating mass-squared asymmetries generate a contribution to $\Delta P$ that is functionally degenerate with a shift in $\delta_{\rm CP}$, inducing an effective, energy-dependent phase $\phi_{\rm eff}(E)$ that causes a CPT-conserving analysis to infer a biased CP phase. 
Using the DUNE experiment as a concrete example, we show that current experimental bounds on $\delta\Delta m^2_{31}$ are already sufficient to produce significant biases in the reconstructed $\delta_{\rm CP}$ and, in CP-conserving scenarios, to fake a near-maximal CP-violation signal. 
We then show how atmospheric neutrino telescopes---specifically IceCube-DeepCore and KM3NeT/ORCA---can break this degeneracy by providing independent, high-statistics constraints on $\delta\Delta m^2_{31}$, and we project the combined sensitivity of present and future configurations as a function of time.

The rest of this article is structured as follows. Section~\ref{sec:pheno} develops the analytic framework, decomposing $\Delta P$ into CP, CPT, and matter contributions and deriving the effective phase shift $\phi_{\rm eff}(E)$ that quantifies the CP--CPT degeneracy. Section~\ref{sec:dune} presents the DUNE sensitivity analysis using \texttt{GLoBES}, illustrating the bias on $\delta_{\rm CP}$ as a function of $\delta\Delta m^2_{31}$, describes the atmospheric neutrino analysis with the \texttt{Pynu} framework~\cite{Arguelles2023MeasuringNeutrinos}, and presents current constraints and future projections from IceCube-DeepCore and ORCA. 
We conclude in Section~\ref{sec:conclusions}.

%% file: Sections/Pheno.tex
\subsection{Decomposition of $\Delta P$}
\begin{figure}[h]
    \centering
    \includegraphics[width=\textwidth]{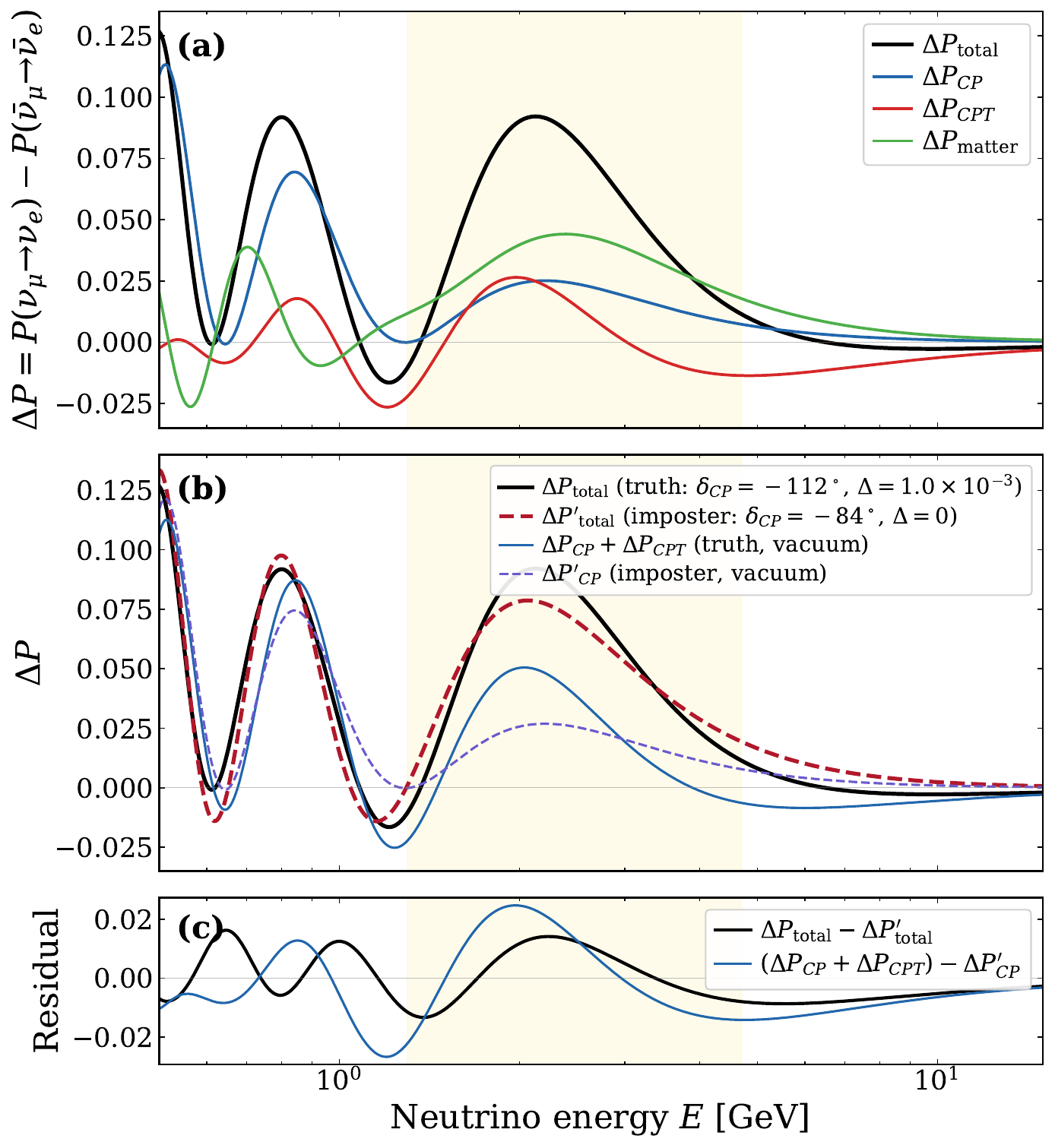}
    \caption{CP--CPT degeneracy at DUNE. \textbf{Panel (a)}~Three-way decomposition of $\Delta P$at the true parameters. \textbf{Panel (b)}~Comparison of the total $\Delta P$ at truth with the impostor hypothesis (dashed red).The vacuum components ($\Delta P_{CP} + \Delta P_{CPT}$ at truth vs.\ $\Delta P'_{CP}$at the impostor) are shown for comparison (blue solid vs.\ purple dashed). \textbf{Panel (c)}~Residual differences: the total residual (black) is $\lesssim 1.4\%$ in the DUNE $\nu_e$ appearance window (1.3--4.7~GeV, gold shading), confirming the approximate degeneracy. All probabilities computed with nuSQuIDS (3-flavor, constant-density matter, $L = 1285$~km, $\rho = 2.848~\text{g/cm}^3$).}
    \label{fig:dune-3-decomp}
\end{figure}
The $\nu_e$ and $\bar\nu_e$ appearance probabilities, in the presence of a mass-squared-splitting asymmetry $\delta \Delta \equiv \Delta_{\nu\bar\nu}(\Delta m^2) = \bar{\Delta} m^2 - \Delta m^2$, are given by the Cervera formula~\cite{Cervera:2000kp} as
\begin{equation}
    \begin{aligned}
            P(\nu_\mu \rightarrow \nu_e) = \alpha^2 + \beta^2  + 2\alpha\beta \cos(\Delta_{31} + \delta_{\textrm{CP}}), \\
            P(\bar\nu_\mu \rightarrow \bar\nu_e) = \bar\alpha^2 + \beta^2  + 2\bar\alpha\beta \cos(\bar\Delta_{31} - \delta_{\textrm{CP}}).
    \end{aligned}
\end{equation}
Here, $\Delta_{ij} \equiv \Delta m^2_{ij} L / (4E)$ is the standard oscillation phase. Following similar conventions as used in~\cite{Nunokawa:2007qh},
the amplitudes $\alpha$, $\bar\alpha$, and $\beta$ are defined by
\begin{equation}
    \begin{aligned}
        \alpha = \sqrt{P_{\rm{atm}}} &= s_{23}\sin2 \theta_{13}\frac{\sin((1-\hat A)\Delta_{31})}{1-\hat A} ,\\
        \bar \alpha = \sqrt{\bar P^{\cancel{\rm{CPT}}}_{\rm{atm}}}  &= s_{23}\sin2 \theta_{13}\frac{\sin((1+ \hat{\bar A})\bar\Delta_{31})}{1+\hat{ \bar A}} ,\\
        \beta = \sqrt{P_{\rm{sol}}}  &= c_{23}\sin2 \theta_{12}\frac{\sin(\hat A \Delta_{21})}{\hat A}.
    \end{aligned}
\end{equation}
where $A$ is the matter potential, and we have introduced the dimensionless ratios $\hat A \equiv A/\Delta$ and $\hat{\bar A} \equiv A/\bar\Delta$, with $\Delta$ ($\bar\Delta$) the mass-squared splitting appropriate to the channel.

Here, the parameters that are related to CPT violation include $\bar \alpha_{\rm{CPT}}$, $\hat{\bar A}$, and $\bar \Delta$.
Expanding $\cos(\Delta \pm \delta_{\rm{CP}})$, we can obtain the difference in neutrino and antineutrino oscillation probabilities
\begin{equation}\label{eq:decomp}
\begin{aligned}
        \Delta P = &(\alpha^2 - \abar^2) \\
        &+ 2\beta \cos \delta_{\rm{CP}}(\alpha \cos \Delta  - \bar\alpha \cos \bar \Delta ) \\
        &- 2\beta \sin \delta_{\rm{CP}} (\alpha \sin \Delta  + \bar \alpha \sin \bar \Delta ).
\end{aligned}
\end{equation}

Observe that the second term is CP-even, and the third term is CP-odd: in vacuum, without CPT violation, and with $\alpha = \bar \alpha$, this term is proportional to the Jarlskog invariant.
However, due to interference terms, it is hard to disentangle the effects of CP, CPT, and the matter potential. 
To disentangle these contributions, denoting the difference in atmospheric oscillation probability difference with and without CPT conservation as $\bar \alpha_{\rm{CPT}} - \bar \alpha_{\cancel{\rm{CPT}}}$, we can define the CPT contribution $\Delta P_{\rm{CPT}}$ as
\begin{equation}
\begin{aligned}
        \Delta P_{\rm{CPT}} = &(\bar \alpha_{\rm{CPT}} - \bar \alpha_{\cancel{\rm{CPT}}}) \\
        + &(\sin \delta_{\rm{CP}} + \cos \delta_{\rm{CP}})\\
     \times&(2 (\delta \bar \alpha) \beta \cos \bar \Delta + 2 \bar \alpha \beta (\cos \Delta - \cos \bar \Delta)).
\end{aligned}
\end{equation}

One identifies the matter-CPT interference term as well as the matter-CP-CPT interference terms, and verifies that $\Delta P_{\rm{CPT}} = 0$ with $\bar \Delta = \Delta$ and $\delta \bar \alpha = 0$. 
With this parameterization, we obtain back the decomposition from Ref.~\cite{Nunokawa:2007qh}
\begin{equation}
    \Delta P = \Delta P_{\rm{CP, matter}} + \Delta P_{\rm{CPT}}.
\end{equation}

For simplicity, we can define $\Delta P_{\rm{CP}} = \Delta P_{\rm{CP}}^{\rm{vacuum}}$ to be the vacuum genuine CP violation, and let $\Delta P_{\rm{matter}} = \Delta P - \Delta P_{\rm{CP}}$ absorb the leftover probability difference in CPT-conserving scenario. 
Another choice, as done in Ref.~\cite{Nunokawa:2007qh}, would be to disentangle matter and CP contributions by defining $\Delta P_{\rm{matter}} = \Delta P (\delta_{\rm{CP}} = \pi /2) - \Delta P (\delta_{\rm{CP}} = 0)$.

We use the \texttt{nuSQuIDS} package~\cite{Arguelles:2022hrt} in a constant-matter-density configuration, taking the DUNE baseline and matter density as our reference accelerator setup, and choose a CPT-violating mass-squared splitting of $\delta \Delta = 1.0 \times 10^{-3}~\text{eV}^2$.
Figure~\ref{fig:dune-3-decomp}(a) shows the resulting decomposition of the total $\Delta P$ into its three components. Panels (b) and (c) further demonstrate that, at a true $\delta_{\rm{CP}} = -112^\circ$, the CP and CPT contributions to $\Delta P$ conspire to produce a
degeneracy: an impostor CPT-conserving solution at $\delta_{\rm{CP}} = -84^\circ$, $\delta\Delta = 0$ that reproduces the true $\Delta P$.

This example of an impostor solution leads us to understand qualitatively first the physical implications of the decomposition of $\Delta P$ in Eq.~\ref{eq:decomp}: CPT violation can effectively \emph{mimic CP violation} in long-baseline experiments.
Two features are crucial:
\begin{enumerate}
    \item CPT-violating terms interfere with CP-sensitive contributions, producing  
    an energy-dependent effect functionally similar to genuine CP violation in vacuum.
    \item Matter effects further modify the interference, enhancing the degeneracy  
    between CP and CPT effects.
\end{enumerate}
This interplay allows for situations in which a true CP phase with small CPT violation   produces the same $\Delta P$ as a CPT-conserving scenario with a shifted CP phase.  

Additional considerations to break or mitigate this degeneracy include:
\begin{itemize}
    \item \textbf{Energy dependence:} The mimicking effect varies with neutrino energy,  
    so combining spectral information across the energy range can help distinguish CP and CPT contributions.
    \item \textbf{Channel complementarity:} Using multiple oscillation channels, such as $\nu_\mu$ disappearance or $\nu_e$ appearance at different baselines, provides independent constraints on CPT-violating parameters.
    \item \textbf{Baseline dependence:} Different experiments experience different matter effects, so combining long-baseline setups improves sensitivity to genuine CP violation.
\end{itemize}
In summary, \emph{an observed CP-odd asymmetry in $\nu_e$ appearance does not  
necessarily indicate genuine CP violation in vacuum}. 
Only by combining multiple channels, baselines, and energy spectra, or by imposing precise constraints on CPT-violating parameters, can this degeneracy be resolved.  
The decomposition of $\Delta P$ into CP, matter, and CPT contributions provides a clear framework to quantify the extent to which CPT can mimic CP violation.

\subsection{Effective CP Phase Shift from CPT Violation}

Now, we turn to a quantitative study of this degeneracy by further parameterizing the decomposition in Eq.~\ref{eq:decomp} through a coordinate transformation.
We start from the CP-dependent part of the probability difference,
\begin{equation}
    \Delta P_{\rm int}(E)
    =
    A(E)\cos\delta_{\rm CP}
    +
    B(E)\sin\delta_{\rm CP},
\end{equation}
where
\begin{align}
    A(E) &= 2\beta \left(\alpha \cos\Delta - \bar\alpha\cos\bar\Delta \right), \\
    B(E) &= -2\beta \left(\alpha \sin\Delta + \bar\alpha\sin\bar\Delta \right),
\end{align}
and recast this expression in the form
\begin{equation}
    \Delta P_{\rm int}(E)
    =
    R(E)\,\sin\big(\delta_{\rm CP} + \phi_{\rm eff}(E)\big),
\end{equation}
with amplitude $R(E) = \sqrt{A^2(E) + B^2(E)}$.
The effective phase shift $\phi_{\rm eff}(E)$ is defined through
\begin{equation}
    \tan\phi_{\rm eff}(E)
    =
    \frac{A(E)}{B(E)},
\end{equation}
which yields
\begin{equation}
    \phi_{\rm eff}(E)
    =
    \arctan\!\left[
    \frac{\alpha \cos\Delta - \bar\alpha\cos\bar\Delta}
    {-\alpha \sin\Delta - \bar\alpha\sin\bar\Delta}
    \right].
    \label{eq:phieff}
\end{equation}

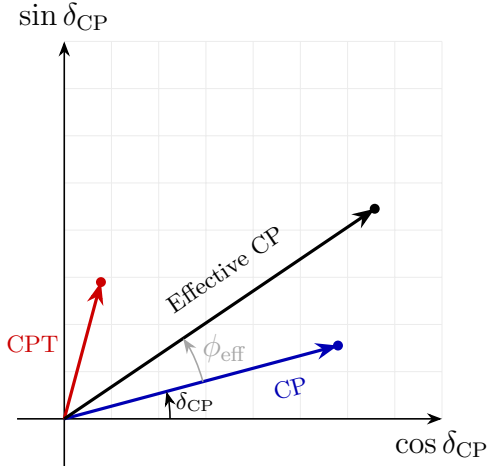
\begin{figure}[t]
    \centering
    \begin{tikzpicture}[
        >=Stealth,
        scale=1.25,
        font=\small
    ]

    % Angle for CP vector
    \def\rotateAngle{15}

    % Coordinates
    \coordinate (O)   at (0,0);
    \coordinate (X)   at (1,0);
    \coordinate (CP)  at ({3*cos(\rotateAngle)},{3*sin(\rotateAngle)});
    \coordinate (CPT) at ({1.5*cos(\rotateAngle+60)},{1.5*sin(\rotateAngle+60)});
    \coordinate (Eff) at ($(CP)+(CPT)$);

    % Light grid
    \draw[gray!15, very thin, step=0.5] (0,0) grid (4,4);

    % Axes
    \draw[->, semithick] (-0.5,0) -- (4,0)
        node[below =1pt] {\large $\cos\delta_{\rm CP}$};
    \draw[->, semithick] (0,-0.5) -- (0,4)
        node[above =1pt] {\large $\sin\delta_{\rm CP}$};

    % Vectors
    \draw[->, very thick, blue!70!black] (O) -- (CP)
        node[pos=0.80, below, sloped, yshift=-4pt] {CP};

    \draw[->, very thick, red!80!black] (O) -- (CPT)
        node[pos=0.55, left=6pt] {CPT};

    \draw[->, very thick, black] (O) -- (Eff)
        node[pos=0.58, above, sloped, yshift=6pt] {Effective CP};

    % End points
    \fill[blue!70!black] (CP) circle (1.5pt);
    \fill[red!80!black]  (CPT) circle (1.5pt);
    \fill[black]         (Eff) circle (1.5pt);

    % Angle delta_CP
    \pic[
        draw,
        ->,
        semithick,
        angle radius=1.4cm,
        "$\delta_{\rm CP}$",
        angle eccentricity=1.25
    ] {angle = X--O--CP};

    % Angle phi_eff
    \pic[
        draw,
        ->,
        semithick,
        gray!70,
        angle radius=1.9cm
    ] {angle = CP--O--Eff};

    \node[gray!70] at ($(O)+(1.7,0.75)$) {\large$\phi_{\rm eff}$};

    \end{tikzpicture}
    \caption{Illustration of the CP, CPT, and effective CP vectors in the
    $(\cos\delta_{\rm CP},\,\sin\delta_{\rm CP})$ plane. The addition of CPT violation leads to an effective shift to the CP phase.}
    \label{fig:tikz}
\end{figure}

This form makes explicit that CPT violation modifies the CP-phase dependence by
introducing an \emph{effective, energy-dependent phase shift}, where we show an illustrative example in Fig.~\ref{fig:tikz}. See Section~\ref{sec:IceCube} for a description of the energy range in which these effects are most significant. 

We can further make this more transparent by considering an expansion of the CPT-violating quantities and showing the first-order contributions: we can denote
\begin{equation}
    \bar\Delta = \Delta + \delta\Delta,
    \qquad
    \bar\alpha_{\rm CPT} = \alpha + \delta\bar\alpha,
\end{equation}

To first order,

\begin{equation}
\begin{aligned}
        \phi_{\rm eff}(E)
    \simeq
    -\frac{1}{2}\delta \Delta_{31}(E)
    &+
    \frac{1}{2}\frac{\delta\bar\alpha(E)}{\alpha(E)}\,\cot\Delta_{31}(E).
\end{aligned}
\end{equation}

This result provides a direct physical interpretation of the CP--CPT degeneracy:
CPT violation effectively induces a shift
\begin{equation}
    \delta_{\rm CP}
    \;\longrightarrow\;
    \delta_{\rm CP} + \phi_{\rm eff}(E),
\end{equation}
with $\phi_{\rm eff}(E)$ depending nontrivially on energy through both the oscillation
phase and matter effects. The term proportional to $\delta\bar\alpha/\alpha$ encodes amplitude-level
CPT violation and matter--CPT interference. The term proportional to $\delta\Delta$ represents a genuine phase shift arising from differences in the mass-squared splittings.

Because $\phi_{\rm eff}(E)$ is energy dependent, a CPT-violating scenario can
closely reproduce the oscillation probability of a CPT-conserving theory with a
shifted $\delta_{\rm CP}$ at a fixed energy, while deviations emerge once spectral
information is taken into account. Supposing that only monochromatic measurements of the CP phase are taken at two energies $E_1$ and $E_2$, then at first order,

\begin{equation}
    \Delta \delta_{\rm CP}^{\rm exp}
    \simeq
    \phi_{\rm eff}(E_2) - \phi_{\rm eff}(E_1).
\end{equation}
Near the first oscillation maximum ($\Delta \sim \pi/2$), this reduces to
\begin{equation}
    \Delta \delta_{\rm CP}^{\rm exp}
    \sim
    - \frac{1}{2}\delta\Delta_{31}(E)
\end{equation}

On the one hand, percent-level shifts in the inferred CP phase can arise from $\mathcal{O}(10^{-3}\,\text{eV}^2)$ differences in the atmospheric mass-squared splitting or comparable fractional changes in the oscillation amplitude (while such effects are constrained by existing oscillation data, this estimate demonstrates that CPT violation at a level not yet excluded could induce experiment-dependent shifts in $\delta_{\rm CP}$); and on the other hand, $\phieff(E)$ being an energy-dependent quantity suggests that different long-baseline experiments probing distinct $L/E$ regimes may infer different effective values of $\delta_{\rm CP}$.

It is worth noting here that the expansion is only valid when $|\delta\Delta| \ll 1$ and $|\delta\bar\alpha| \ll \alpha$; however, the expansion is qualitatively valid in showing the energy dependence of the $\phieff(E)$ term, and the full prescription (Eq.~\ref{eq:phieff}) is always valid.

%% file: Sections/Exp.tex
\begin{figure*}[th!]
    \centering
    \includegraphics[width=\linewidth]{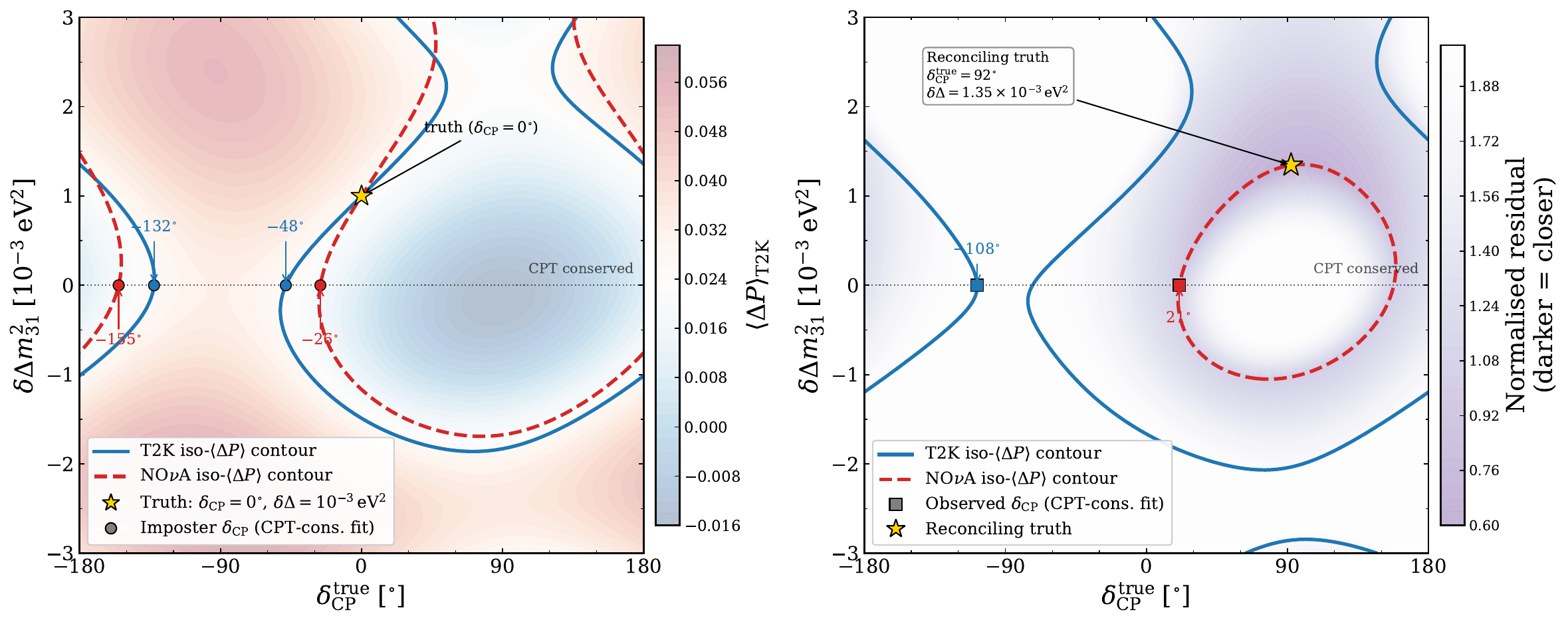}
    \caption{
    CP--CPT degeneracy manifolds for T2K (295~km, solid blue) and NOvA (810~km,
    dashed red) in the
    $(\delta_{\rm CP}^{\rm true},\;\delta\Delta m^2_{31})$ truth plane.
    The observable is the flux-weighted CP asymmetry
    $\langle\Delta P\rangle$,
    computed via the Cervera analytic formula with matter effects. Fluxes are assumed to be Gaussian distributed around respective characteristic fluxes.
    \textbf{Left panel (image):}
    iso-$\langle\Delta P\rangle$ contours passing through a CPT-violating
    truth at $\delta_{\rm CP} = 0^\circ$,
    $\delta\Delta m^2_{31} = 10^{-3}$~eV$^2$ (gold star).
    Where each contour crosses the CPT-conserving axis ($\delta\Delta m^2_{31} = 0$),
    the corresponding ``impostor'' $\delta_{\rm CP}$ is the value a
    CPT-conserving fit would recover.
    The background colour field shows the T2K $\langle\Delta P\rangle$ value.
    \textbf{Right panel (preimage):}
    given T2K's observed $\delta_{\rm CP} = -108^\circ$ and NOvA's observed
    $\delta_{\rm CP} = +21^\circ$ under CPT conservation, the preimage curves
    trace the locus of CPT-violating truth points that would produce each
    measurement.
    Their intersection (gold star) is the unique CPT-violating truth
    ($\delta_{\rm CP}^{\rm true} \approx 109^\circ$,
    $\delta\Delta m^2_{31} \approx 1.48 \times 10^{-3}$~eV$^2$)
    that reconciles both experiments simultaneously.
    The background colour field shows the normalised residual (darker regions
    are closer to satisfying both constraints). 
    }
    \label{fig:manifold}
\end{figure*}

\subsection{Accounting for tension in NOvA-T2K}

The T2K and NOvA collaborations have performed a joint fit on neutrino oscillation parameters, and presented a mild tension in their measurements of the CP-violating phase~\cite{T2K:2025wet}.
Our parameterization of the CPT violation in oscillations provides a natural mechanism and explanation to account for the difference in best-fit points $\delta_{\rm CP}$ between T2K and NO$\nu$A.
Here, we provide viewpoints from two directions as shown in Fig.~\ref{fig:manifold}, where we show that the true $\Delta P_{\nu \bar \nu}$ is equivalence classes forming submanifolds in the $\dCP^{\rm{true}}$ and $\delta\Delta^{\rm{true}}$ plane. 
The physical observable is 
\begin{equation}
    \langle\Delta P\rangle
      = \langle P(\nu_\mu\!\to\!\nu_e)\rangle_{\Phi_\nu}
      - \langle P(\bar\nu_\mu\!\to\!\bar\nu_e)\rangle_{\Phi_{\bar\nu}}.
\end{equation}
For simplicity, we also assume that the flux is Gaussian distributed around the characteristic flux of the two experiments with spread mimicking the on- and off-axis effects: T2K assumes $\mu_E = 0.6$ GeV with $\sigma_E = 0.1$ GeV, and NOvA assumes $\mu_E = 1.9$ GeV with $\sigma_E = 0.45$ GeV.

On the one hand,  we assume a CPT-violating but CP-conserving truth, and look at the equivalent points in the submanifold where the physical observable $\Delta P$ is invariant, and we find that they cross the CPT-conserving axis at different points due to the different baselines. 
The key observation is that any point along this submanifold is perfectly equivalent to any other point, so there's nothing special about these labeled CPT-conserving solutions.
On the other hand, we ask the preimage question: given the best-fit points of the NOvA and T2K experiments, respectively, where the $\Delta P$ is computed with the CPT-conserving hypothesis, what are the equivalence manifolds in the $(\delta \Delta, \, \dCP)$ plane of the two experiments? 
We can even further trace the locus on these points to find a best tension-reconciling CPT-non-conserving truth. 
One can see that the normalized difference in $\Delta P$ across the flux range reduces by 75\% if we allow for a CPT-non-conserving truth.
The mere existence of the distinct intersection points constitutes a CPT-based explanation for the T2K–NOvA tension that is invisible to any CPT-conserving analysis, confirmed by the proximity of the two preimage contours.

The two experiments probe the same $L/E$ regime, but with neutrino beams that span distinct baselines and peak at different energies, implying a varying impact of matter effects in each case. This leads to separate degeneracy manifolds in the $(\delta_{\rm CP}, \delta \Delta m^2_{31})$ plane, whose CPT-conserving intersections are generically displaced from one another.
The degeneracy is exact at the level of the flux-averaged asymmetry $\langle\Delta P\rangle$ but approximate in the spectral sense, since $\phi_{\rm eff}(E)$ varies across the flux window; this is the handle that future high-statistics experiments can exploit to break it. 

Even with the flux information $\phi(E)$ and reconstruction, a single accelerator experiment with a fixed matter effect profile and some energy spectral information typically still suffers from the CP-CPT degeneracy if a fit for oscillation parameters is performed in both horn currents combined. 
Supp.~Fig.~\ref{fig:nova_combines} in the Appendix shows the CP phase measurements for NOvA following the treatment discussed here. In the next subsection, we also show a more detailed discussion for DUNE, where we further include flux information and detector energy reconstruction effects. 
We find in these plots that the measured $\dCP$ values follow the $\Delta P$ isocontours approximately, validating its usage as an effective observable at truth level.

\begin{figure*}[t]
    \centering
    \includegraphics[width=\textwidth]{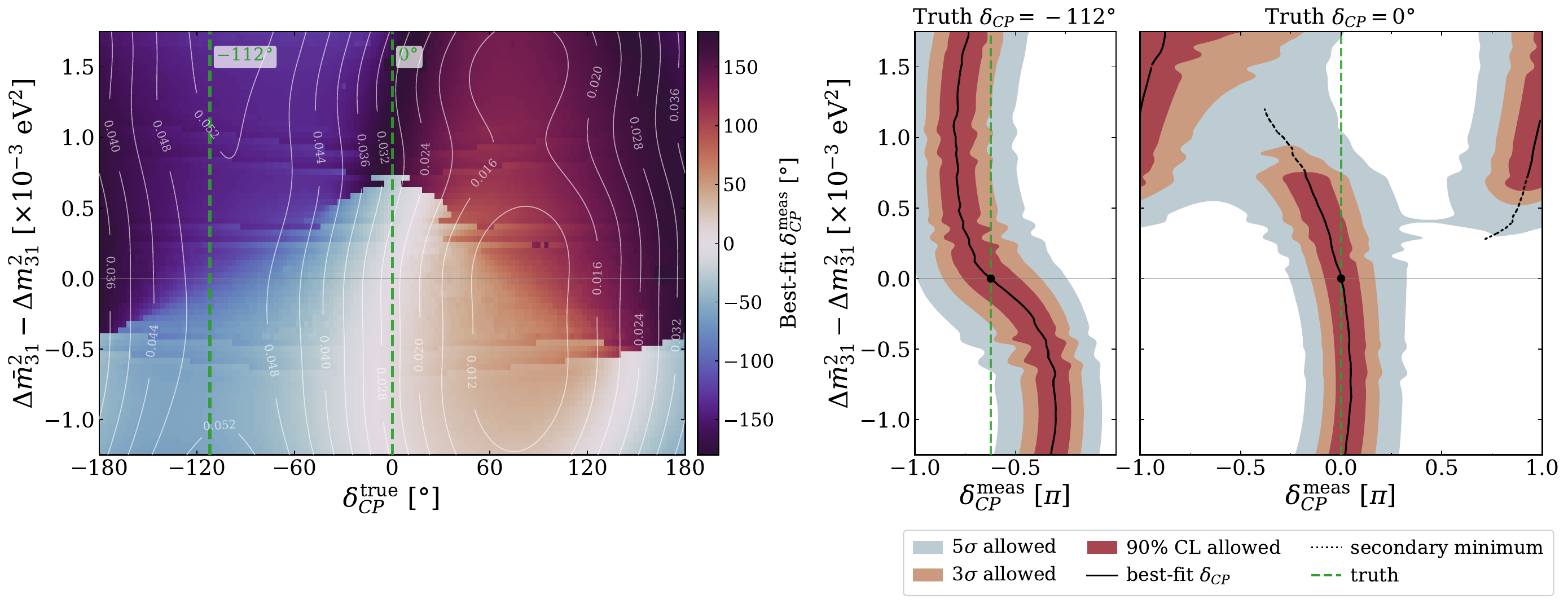}
    \caption{CPT-induced bias on $\delta_{CP}$ at DUNE (40~kt LAr, 1.2~MW, 7~yr staged
    FHC+RHC).
    \textbf{Left panel:} best-fit $\delta_{CP}^{\rm meas}$ obtained from a CPT-conserving fit across the truth plane
    $(\delta_{CP}^{\rm true},\;\delta\Delta m^2_{31})$,
    where $\delta\Delta m^2_{31}\equiv\Delta\bar{m}^2_{31}-\Delta m^2_{31}$.
    White contours show the flux-weighted CP asymmetry
    $\langle\Delta P\rangle
     =\langle P(\nu_\mu\!\to\!\nu_e)\rangle_{\Phi_\nu}
     -\langle P(\bar\nu_\mu\!\to\!\bar\nu_e)\rangle_{\Phi_{\bar\nu}}$
    computed with the actual DUNE FD beam flux (separate FHC $\nu_\mu$ and
    RHC $\bar\nu_\mu$ weights).
    Green dashed lines mark the two truth $\delta_{CP}$ values ($-112^\circ$ and
    $0^\circ$) whose vertical slices are expanded in the right panels.
    \textbf{Right panels:} confidence bands for the measured $\delta_{CP}$ as a
    function of $\delta\Delta m^2_{31}$, using 8-rule GLoBES implementation  profiling over
    $\theta_{23}$, $\Delta m^2_{31}$, $\theta_{13}$. A basin switch near $\delta\Delta m^2_{31}\approx+0.76\times10^{-3}$~eV$^2$
    is visible in the right panel, creating an island where
    undetected CPT violation produces a false measurement of near-maximal
    CP violation. }
\label{fig:band}
    
\end{figure*}
This result motivates the central question of the remainder of the paper: \textbf{can an independent, external constraint on $\delta \Delta m^2_{31}$ from atmospheric neutrinos break this degeneracy before DUNE begins its CP measurement?}

\subsection{CP-CPT degeneracy in DUNE}

We first present the problem of the degeneracy between CP and CPT in accelerators, using the DUNE experiment as an example.
We use the GLoBES package with the DUNE configuration~\cite{Huber_2005,Huber_2007}. 
In this fit, we assume 7.5 years of FHC (Forward Horn Current) and RHC (Reverse Horn Current) equally, but we assume that only a combined fit will be performed, assuming that CPT is conserved, where $\delta\Delta\equiv \Delta\bar{m}^2_{31} - \Delta m^2_{31} = 0$. 

CPT violation in the context of DUNE has a significant impact on the measurement not only due to the combination of FHC and RHC in analysis $R_{\nu_e} = R_{\nu_e}^{\rm{FHC}} + R_{\nu_e}^{\rm{RHC}}$, but also due to the wrong sign contamination of the beam.
This is because, taking $\nu_e$ appearance rate in RHC as an example, the event rate is computed by the sum of $\nu_\mu$ and $\bar \nu_\mu$ components
\begin{equation}
\begin{aligned}
        R^{\rm{RHC}}_{\nu_e, \bar\nu_e} &= R^{\rm{RHC, wrong}}_{\nu_\mu \rightarrow \nu_e} + R^{\rm{RHC}}_{\bar \nu_\mu \rightarrow \bar \nu_e}, \\
        &=\phi^{\rm{RHC}}_{\nu_\mu} (E) \cdot P_{\mu e} (E; \Delta_{31}) \cdot \sigma ^{\rm{CC}}_{\nu_e}(E) \cdot \epsilon_{\rm{det}}(E) \\
        &+\phi^{\rm{RHC}}_{\bar \nu_\mu} (E) \cdot \bar P_{\bar \mu \bar e} (E; \bar \Delta_{31}) \cdot \sigma ^{\rm{CC}}_{\bar \nu_e}(E) \cdot \epsilon_{\rm{det}}(E),
\end{aligned}
\label{eq:DUNE-rate}
\end{equation}
where the first contribution in flux alone is $15\%$~\cite{dunecollaboration2016longbaselineneutrinofacilitylbnf}, which can correspond to $35\%$ in event rate due to oscillation and cross section. 
With CPT violation, the neutrino and antineutrino oscillation probabilities are different, leading to different distortions of the event distribution that hide in plain sight and can be mistaken for a CP-violation effect.

In GLoBES, the calculation of event rates and therefore likelihood for a set of oscillation parameters is obtained via the usage of ``rules": an analysis-level reconstructed spectrum. Conventionally, such as shipped in the DUNE configuration~\cite{Huber_2005,Huber_2007}, there are four such spectra: FHC appearance, RHC appearance, FHC disappearance, and RHC disappearance. However, each of these spectra, as shown in~\ref{eq:DUNE-rate}, contains both the right and the wrong sign contributions, and assigning a single $\Delta m^2_{31}$ value cannot account for the effect of CPT violation. We therefore use an equivalent 8-rule configuration in which each physical rule is split into a neutrino-only and an antineutrino-only sub-rule. The CPT-violating Asimov prediction is generated by evaluating the neutrino sub-rules with \(\Delta m^2_{31}\) and the antineutrino sub-rules with \(\Delta\bar m^2_{31}\), thereby realizing Eq.~\eqref{eq:DUNE-rate} at the rate level.
It is worth noting that the 8 sub-rules are not treated as eight independently observed samples. After the neutrino and antineutrino components are generated with their respective mass splittings, the corresponding sub-rules are summed back into the 4 physical DUNE spectra in order to enforce shared signal and background normalization nuisance parameters for each physical spectrum.

The right two panels of Fig.~\ref{fig:band} show two panels assuming a near-maximal CP violation and zero CP violation Asimov truths, respectively.
Varying the Asimov data set true $\delta \Delta$ on a grid, $\delta_{CP}^{\textrm{meas}}$ is scanned over $[-\pi, \pi]$ while profiling over $\theta_{23}$, $\Delta m^2_{31}$, and $\theta_{13}$.
We show that the band of fitted $\delta_{CP}^{\textrm{meas}}$ assuming CPT conservation shifts from the true value. 
In the case of existing CP-violation with $\delta_{\rm{CP}} = -112 \degree$, assuming $\delta \Delta \leq -0.3 \times 10^{-3}$~eV$^2$, which is well below current bounds of $|\delta \Delta| \leq 0.8 \times 10^{-3}$~eV$^2$~\cite{Barenboim_2018}, the allowed value of $\delta_{\rm{CP}}$ incorrectly assuming CPT conservation completely misses the true value of CP violation. 
On the other hand, if the nature were CP-conserving, a CPT violation of $\delta \Delta \geq 0.8 \times 10^{-3}$~eV$^2$ could lead to double-minima structure in the $\chi^2$ profile of the $\delta_{\rm{CP}}$ fit when the true $\dCP$ is close to 0, leading to a potential false detection of near-maximal CP violation.

\begin{figure}
    \centering
    \includegraphics[width=\linewidth]{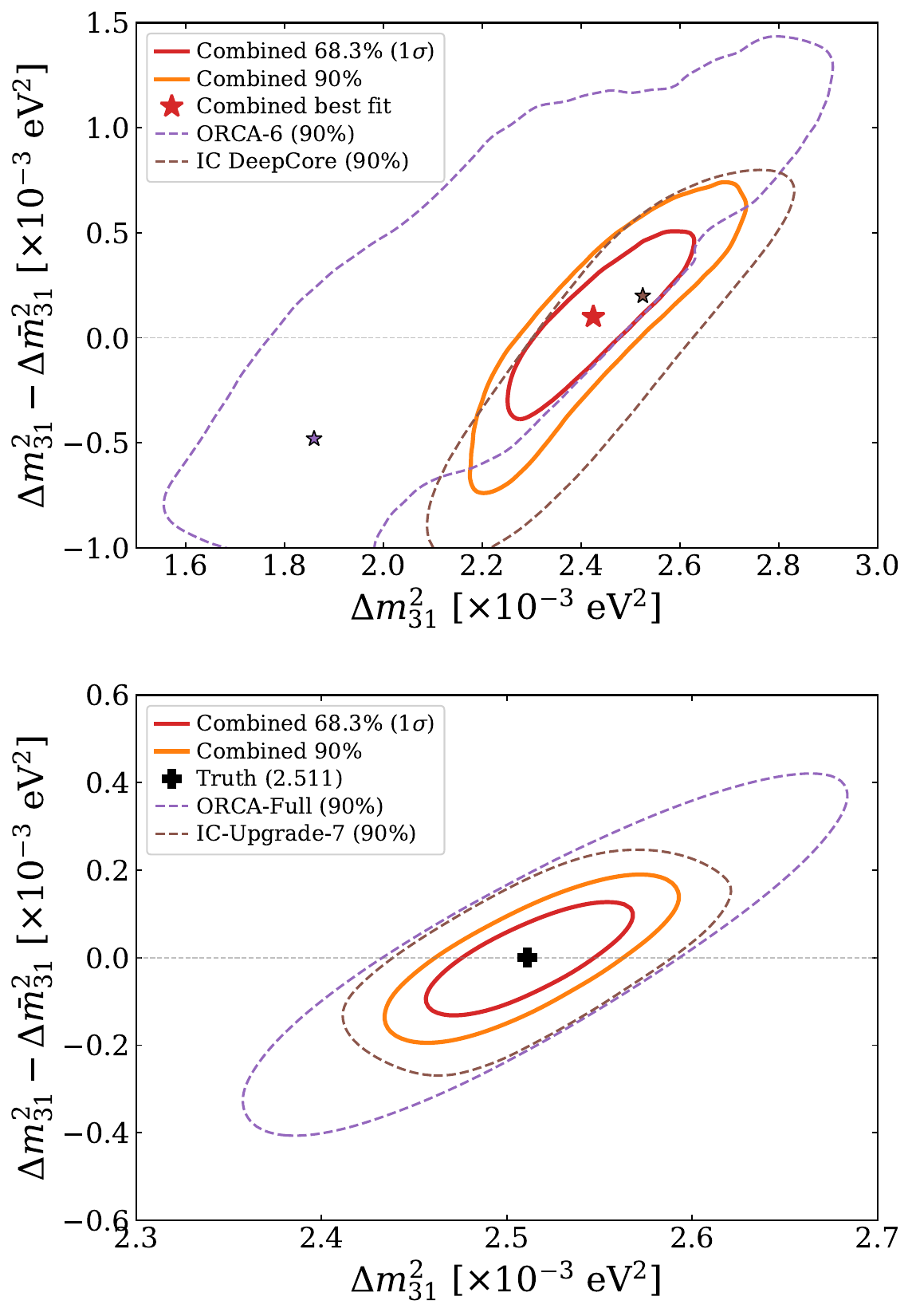}
    \caption{Two-dimensional confidence contours in the $(\Delta m^2_{31},\; \delta\Delta m^2)$ plane.
    \textbf{Top:} Data fit using ORCA 433 kt-yr and IceCube DeepCore 7.74-year data.
    Stars mark the best-fit points for each experiment and their combination.
    \textbf{Bottom:} Asimov sensitivity projection for ORCA-Full (5~yr) and IceCube Upgrade (7~strings configuration, 10~yr livetime).
    The cross marker indicates the injected truth at $\Delta m^2_{31} = 2.511 \times 10^{-3}~\mathrm{eV}^2$, $\Delta = 0$, assuming Normal Ordering.
    Contours correspond to 68.3\% ($1\sigma$) and 90\% confidence levels for two degrees of freedom.
    }
    \label{fig:combined-fit}
\end{figure}

This double minima and basin switch behavior can be understood intuitively as two competing fit minima. 
For neutrinos and antineutrinos respectively, the $\dCP$ term enters in the interference probability terms in $+$ and $-$ signs, i.e.,
\begin{equation}
    \begin{aligned}
        \Delta P ^{\rm{int}}_{\nu}(\dCP) &= 2\alpha \beta \cos(\Delta_{31}+\dCP),\\
        \Delta P ^{\rm{int}}_{\bar \nu}(\dCP) &= 2\bar\alpha \bar\beta \cos(\bar\Delta_{31}-\dCP).
    \end{aligned}
\end{equation}
Therefore, when we shift $\dCP \rightarrow \dCP + \pi$, we flip the sign of both interference terms:
\begin{equation}
        \Delta P ^{\rm{int}}_{\nu, \bar \nu}(\dCP + \pi) = -\Delta P ^{\rm{int}}_{\nu, \bar \nu}(\dCP).
\end{equation}
This creates the two competing minima near $\dCP \approx 0 \degree$. 
At small $\delta\Delta$, the fit stays near the true $\dCP$ and absorbs the $\bar \nu$ rate mismatch induced by CPT violation through shifts in $\theta_{23}^{\rm{meas}}$ and $\Delta_{31}^{\rm{meas}}$ in the non-interference $\alpha^2 + \beta^2$ terms, and the interference terms $\Delta P_{\rm{int}}$ remains close to their true value.
However, at larger $\delta\Delta$, such a spectral difference in the $\bar \nu$ channel cannot be absorbed as well by the non-interference terms, and as a result, a competing fit method that better accommodates the enhanced $\bar\nu_e$ appearance rate wins when $\delta\Delta > 0$ at the cost of worsening the $\nu_e$ appearance fit. 
In the bottom right panel of Fig.~\ref{fig:combined-fit}, where the near-truth fit is around $\dCP^{\rm{meas}} \approx -31 \degree$ and flips to $\dCP^{\rm{meas}} \approx +169 \degree$. One can further observe from the top panel that such basin flip occurs at $\delta\Delta>0$ for $\dCP \approx 0 \degree$ and at $\delta\Delta<0$ for $\dCP \approx 180 \degree$. 
Furthermore, one can see that in each region across the discontinuity, the $\dCP^{\rm{meas}}$ roughly follows the $\Delta P$ isocontours for analytic calculation. 
Specific event distributions illustrating the two competing fit strategies as favoring the $\nu_e$ and $\bar \nu_e$ channels can be found in the Appendix at Supp.~Fig.~\ref{sppfig:dune_evt_dist}.

The sensitivity of DUNE to neutrino oscillation parameters can also be enhanced through measurements of the atmospheric neutrino flux~\cite{Kelly:2019itm,Ternes:2019sak,DUNE:2020ypp}. In contrast to the beam-based analysis discussed above, atmospheric neutrinos probe a broad range of baselines and energies, with the strongest sensitivity to CP-violating effects arising in the sub-GeV region~\cite{Martinez-Soler:2019nhb}. We investigate how CPT violation modifies the extraction of the CP phase in this complementary channel and find that, when CPT is violated, the value of $\delta_{\textrm{CP}}$ inferred under the assumption of CPT conservation can deviate significantly from the true value, as illustrated in Fig.~\ref{fig:band}. Moreover, we observe that the correlation between $\delta_{CP}^{\rm meas}$ and $\delta(\Delta m^2_{31})$ differs qualitatively from that obtained in the beam analysis, reflecting the distinct energy dependence of CPT-violating effects and the dominance of lower-energy events in the atmospheric sample (see Section~\ref{sec:pheno}).

\subsection{Uncovering the impostor CP with IceCube and ORCA}
\label{sec:IceCube}

To extract meaningful $\delta_{\rm CP}$ measurements from
long-baseline accelerator neutrino experiments without fitting for oscillation parameters in FHC and RHC
separately, we turn to atmospheric neutrino telescopes to place an
independent constraint on $\delta\Delta m^2_{31}$. 
Atmospheric neutrinos offer a qualitative advantage that long-baseline beams cannot replicate: a natural decoupling between the CP-sensitive and CPT-sensitive observables. 
The CP-violating phase $\delta_{\rm CP}$ enters the oscillation probability through the interference between the solar and atmospheric amplitudes, a term suppressed by $\alpha_{21} = \Delta m^2_{21}/\Delta m^2_{31}$ that is only significant below $\sim\!1~\text{GeV}$, where the solar oscillation length is commensurate with trans-Earth baselines. 
Above this
threshold the rapid $\Delta m^2_{21}$-driven oscillations average out
over any realistic energy resolution, effectively decoupling
$\delta_{\rm CP}$ from the multi-GeV sample. The CPT-violating
asymmetry $\delta\Delta m^2_{31}$, by contrast, acts through the
atmospheric channel and is most prominent above $1~\text{GeV}$,
where it shifts the $\bar\nu_\mu$ disappearance minimum relative
to the $\nu_\mu$ minimum, generating a characteristic dipole
pattern in the $(E,\cos\theta_z)$ plane (Supp.~Figs.~3--6). The
two effects therefore, occupy largely distinct regions of the
observable phase space, allowing a combined fit over the full energy
range to constrain $\delta\Delta m^2_{31}$ with little contamination
from the unknown $\delta_{\rm CP}$. An additional handle comes from
the MSW resonance: in normal ordering the resonance enhances
neutrino conversion at $E\sim 3$--$6~\text{GeV}$, while CPT
violation shifts the resonance condition independently in the
antineutrino sector, producing a matter-asymmetric signature
that is qualitatively distinct from any CP effect. The sensitivity
to $\delta\Delta m^2_{31}$ is therefore concentrated in the
antineutrino-driven disappearance sample for normal ordering and
in the neutrino sample for inverted ordering, and the inclusive
atmospheric flux — containing both neutrinos and antineutrinos
across a wide range of energies and baselines — exploits both
simultaneously in a single combined fit.

\begin{figure*}
    \centering
    \includegraphics[width=\linewidth]{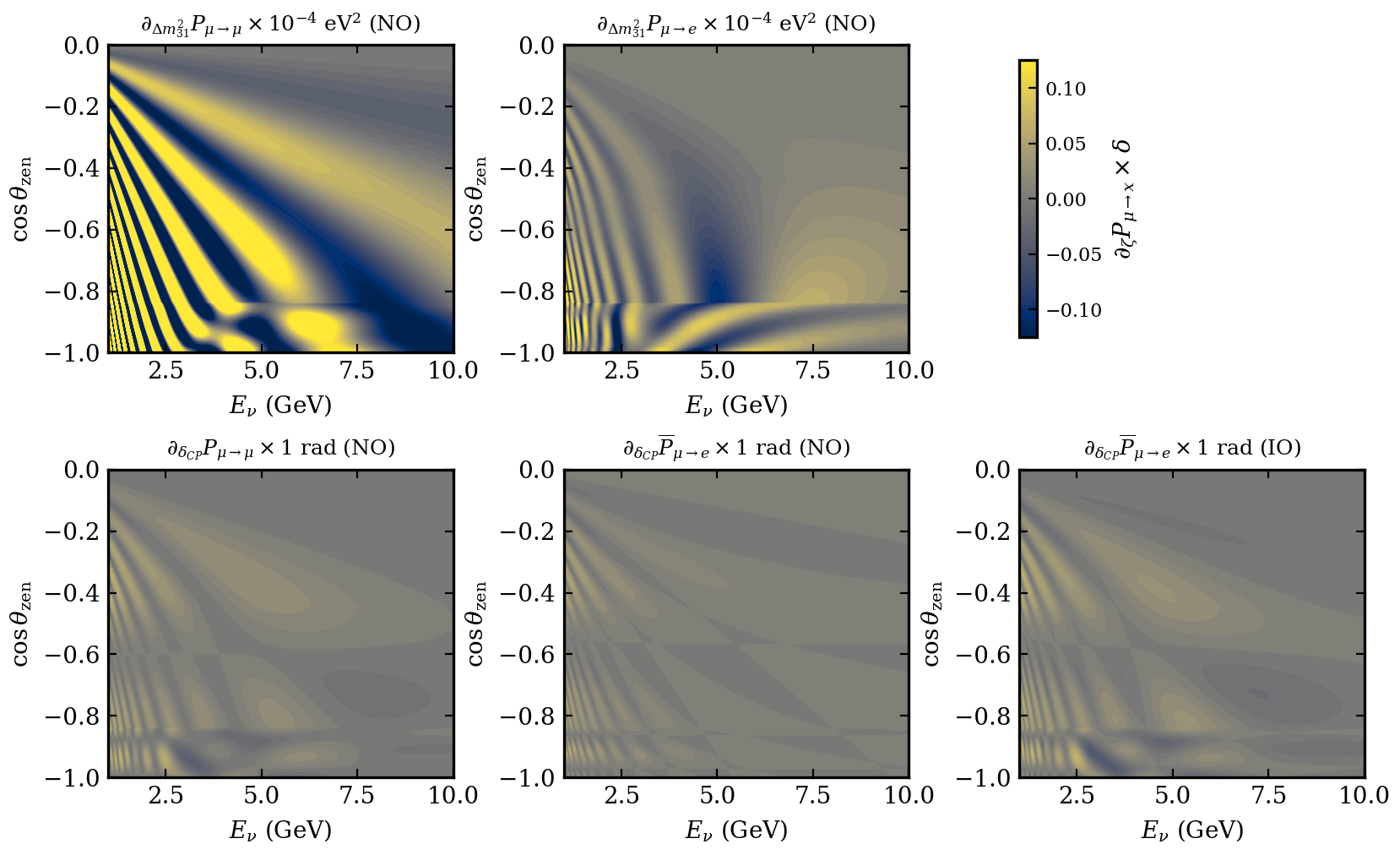}
    \caption{Two-dimensional \textit{oscillograds} as a function of neutrino energy (horizontal axis) and zenith direction (vertical axis).
    \textbf{Top row:} Derivative of the oscillation probabilities with respect to $\Delta m^2_{31}$ for the muon-to-muon (left) and muon-to-electron (center) neutrinos assuming normal ordering (NO).
    \textbf{Bottom row:} Derivative of the oscillation probabilities with respect to $\delta_{CP}$ for the muon-to-muon neutrinos (left) and muon-to-electron antineutrinos (center) assuming normal ordering (NO), and muon-to-electron antineutrinos assuming inverted ordering (IO).
    }
    \label{fig:oscgrads}
\end{figure*}

This decoupling is made explicit in Fig.~\ref{fig:oscgrads},
which shows the derivatives $\partial P_{\alpha\beta}/\partial\Delta
m^2_{31}$ and $\partial P_{\alpha\beta}/\partial\delta_{\rm CP}$ in
the $(E,\cos\theta_z)$ plane for the $\nu_\mu$ disappearance and
$\nu_\mu\to\nu_e$ appearance channels, for both neutrinos and
antineutrinos under normal and inverted ordering, using the \texttt{CHIC} code~\cite{10.1088/1361-6471/ae7818, chic}. The
$\partial P_{\mu\mu}/\partial\Delta m^2_{31}$ panel is dominated by
a large localized structure at the MSW resonance ($E\sim6$--$8$ GeV,
$\cos\theta_z\lesssim-0.8$) in normal ordering, while the
$\partial P_{\mu\mu}/\partial\delta_{\rm CP}$ panel is nearly
structureless across the same region. The two parameters are
therefore, effectively orthogonal in the observable space, most
accessible to IceCube and ORCA, providing a clean experimental
handle on $\delta\Delta m^2_{31}$ that is not contaminated by the
unknown value of $\delta_{\rm CP}$.

Using the \texttt{Pynu} framework, introduced in previous sensitivity studies of atmospheric neutrinos to oscillation parameters~\cite{Arguelles:2022hrt,Olavarrieta:2024eaq}, we perform data fits based on existing data release from the IceCube (IceCube-DeepCore) and ORCA (ORCA-6) experiments~\cite{km3netcollaboration2024measurementneutrinooscillationparameters,Abbasi_2023,ORCA-data}. 

For the data fit, we are using 8 years of IceCube DeepCore data, and 433 kt-yr of ORCA-6 data; for the sensitivity projection, we are assuming 10 years of 7-string IceCube Upgrade~\cite{ICupMC}, five of which have been recently deployed, along with 5 years of the full ORCA detector, whose simulation we have developed in previous work~\cite{Arguelles:2022hrt}.
In both cases, the combined results profile over $\sin^2\theta_{23}$ and $\delta_{\rm{CP}}$ in addition to scanning over $\Delta m^2_{31}$ and $\delta \Delta$. The systematic and statistical treatments are partially inherited from previous work~\cite{Arguelles:2022hrt} but with important changes to account for the changes in data releases, which includes hyperplane systematics and Monte Carlo statistics uncertainties. We do not include DUNE or Hyper-Kamiokande in this analysis, as both detectors are expected to collect limited statistics in the multi-GeV region, where the atmospheric neutrino flux is relatively low.

The Monte Carlo datasets are collections of simulated events ${e}$, each carrying a weight $w_e$. For the ORCA release each event also ships with an MC statistical variance $\sigma_e^2$; the IceCube DeepCore release does not provide one, so we adopt the Poisson estimator $\sigma_e^2 = w_e^2$. The MC variance in analysis bin $i$ is then $\sigma_{{\rm MC},i}^2 = \sum_{e\in i}\sigma_e^2$. 
To mimic what is done in the latest ORCA-6 analysis, our test statistics follow the Barlow–Beeston light prescription~\cite{Barlow:1993jj,conway2011incorporatingnuisanceparameterslikelihoods}: bin $i$ is assigned an auxiliary scale factor $\beta_i$ that absorbs its MC statistical uncertainty on the predicted yield. Writing $O_i$ for the observed count, $E_i(\bm\eta)$ for the total expectation (neutrino plus muon background), and $\sigma^2_{{\rm MC},i}$ for the corresponding MC variance, the test statistic minimised at each grid point reads
\begin{widetext}
\begin{equation}
\label{eq:chi2-bb}
\chi^2(\bm\eta)
  = 2\!\sum_{{\rm Exp}}\sum_{i\in{\rm Bins}}\!
        \left[\,\beta_i E_i(\bm\eta) - O_i
              + O_i\,\ln\!\frac{O_i}{\beta_i E_i(\bm\eta)}\right]
  + \sum_{{\rm Exp}}\sum_{i\in{\rm Bins}}\frac{(\beta_i-1)^2}{\tau_i}
  + \sum_{j\in{\rm Syst}}\!\left(\frac{\tilde\eta_j-\eta_j}{\sigma_j}\right)^{\!2},
\end{equation}
\end{widetext}
with $\tau_i \equiv \sigma^2_{{\rm MC},i}/E_i^2(\bm\eta)$. The first sum
is the saturated Poisson term used in our previous analysis, the
second is the penalty term for the departure of $\beta_i$ from unity, and the third
are the penalty terms for Gaussian-modeled nuisance parameters; the mean and spread of the nuisance parameters are given in Supp. Table~\ref{tab:nuisance}.

The optimal $\beta_i$ at each bin minimises Eq.~\eqref{eq:chi2-bb} in
closed form: it is the positive root of a quadratic,
\begin{equation}
\label{eq:beta-closed-form}
\beta_i \;=\; \tfrac{1}{2}\!\left(-b_i + \sqrt{b_i^{\,2}-4 c_i}\right),
\end{equation}
where $b_i = E_i\tau_i - 1$ and $c_i = -O_i\tau_i$. With this minimization $\beta_i$ is eliminated analytically at every evaluation of
$\chi^2$ rather than minimised numerically.

The total per-bin expectation includes an additive atmospheric-muon
background scaled by a per-experiment normalisation nuisance
$\alpha_\mu^{\rm exp}$:
\begin{equation}
\label{eq:muon-add}
\begin{aligned}
        E_i^{\rm exp}(\bm\eta) \; &=\; E_i^{\nu,\,{\rm exp}}(\bm\eta)
                            \;+\; \alpha_\mu^{\rm exp}\, M_i^{\rm exp}, \\
    \sigma^2_{{\rm MC},i} \; &=\; \sigma^2_{\nu,i} + (\alpha_\mu^{\rm exp})^2\,\sigma^2_{\mu,i}.
\end{aligned}
\end{equation}

\begin{figure}
    \centering
    \includegraphics[width=\linewidth]{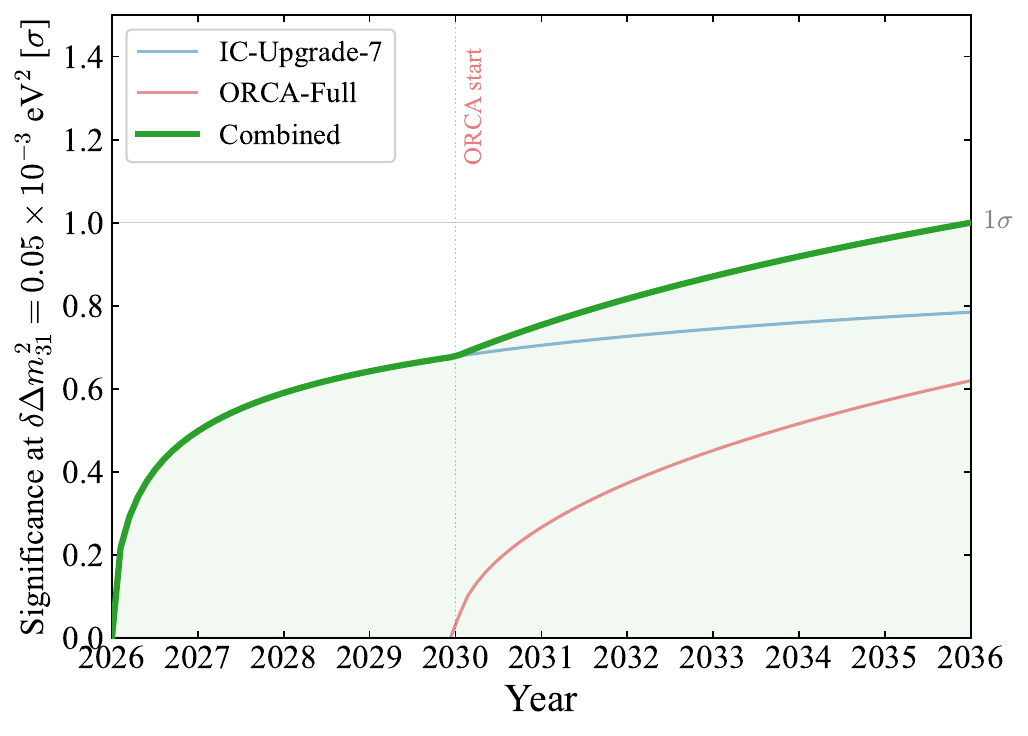}
    \caption{Projected sensitivity to CPT violation as a function of calendar year, expressed as the significance ($\sqrt{\Delta\chi^2}$ in units of $\sigma$) at a fixed CPT asymmetry of $\delta\Delta m^2_{31} = 0.05 \times 10^{-3}~\mathrm{eV}^2$.
    The IceCube Upgrade (blue) begins data-taking in 2026; ORCA-Full (red) is assumed to start in 2030.
    The combined significance (green, with shaded fill) is obtained by adding the individual $\Delta\chi^2$ contributions in quadrature.
    The gray dotted line marks the $1\sigma$ threshold.
}
    \label{fig:chisq-year}
\end{figure}

Using this statistical treatment, we have obtained the confidence regions, shown in Fig.~\ref{fig:combined-fit}. With current data from IceCube and ORCA, we can constrain $| \delta\Delta m^2_{31}|\leq 0.57\times10^{-3}~\text{eV}^2$ (see Suppl. Fig.~\ref{fig:datafit_1d} for 1-dimensional $\chi^2$ profile). However, this is with IceCube DeepCore and ORCA6, where these two low-energy ($\mathcal{O}(1)~\text{GeV}$) experiments will be superseded by their respective successors IceCube Upgrade and the full ORCA detector. 
In the next decade, we will be able to constrain the $\delta \Delta$ CPT violation to be less than $10^{-4}$ eV$^2$ at 90\% confidence level, providing sufficient support to the upcoming DUNE measurement, ruling out the parameter spaces where an impostor solution might exist. 
Assuming that the nature is CPT-conserving, Fig.~\ref{fig:chisq-year} shows the amount of significance we can exclude the current best fit, which is only very slightly biased towards a CPT-violating truth at $5\times10^{-5}$ eV$^2$, as a function of time in years. Starting with the 7-string IceCube-Upgrade configuration, we consider the addition of ORCA full detector in 2030~\cite{KM3NeT_IN2P3_status_2026}, we can rule out this point at 1$\sigma$ in 10 years, effectively ruling out CPT as a source of degeneracy where it is no longer able to produce impostor solutions that differ significantly in 
$\dCP^{\rm{meas}}$.

%% file: Sections/Conclusion.tex
We have studied the interplay between CPT violation and CP violation
in long-baseline neutrino oscillation, with the concrete goal of
understanding whether a CPT-conserving analysis of next-generation
experiments can be trusted to deliver an unambiguous measurement of
$\delta_{\rm CP}$.

The answer, in the absence of external constraints on
$\delta\Delta m^2_{31}$, is no. CPT violation parametrized as an
asymmetry between the neutrino and antineutrino atmospheric
mass-squared splittings induces an effective, energy-dependent phase
shift $\phi_{\rm eff}(E)$ in the $\nu_\mu \to \nu_e$ appearance
asymmetry $\Delta P$. Because $\phi_{\rm eff}(E)$ varies with energy,
it cannot be absorbed into a single redefinition of $\delta_{\rm CP}$
across the full spectrum; nevertheless, once integrated over a
realistic flux window, it produces a flux-averaged $\langle\Delta
P\rangle$ that is indistinguishable from that of a CPT-conserving
scenario with a shifted CP phase. The result is a continuous family of
$(\delta_{\rm CP}, \delta\Delta m^2_{31})$ truth points — a degeneracy
manifold — that are experimentally equivalent at the level of the
observable a single long-baseline experiment can access.

This manifold has direct consequences for the current experimental
landscape. T2K and NO$\nu$A probe different $L/E$ regimes and
therefore inhabit geometrically distinct degeneracy manifolds in the
$(\delta_{\rm CP}, \delta\Delta m^2_{31})$ plane. Their CPT-conserving
best-fit $\delta_{\rm CP}$ values correspond to different intersection
points of their respective manifolds with the $\delta\Delta = 0$ axis.
The tension between those intersections is therefore not a fundamental
disagreement between the experiments — it is the geometric consequence
of fitting a single parameter to data that is sensitive to two. We
showed explicitly that the observed T2K and NO$\nu$A central values
are simultaneously explained by a CPT-violating truth at
$\delta_{\rm CP}^{\rm true} \approx 109^\circ$,
$\delta\Delta m^2_{31} \approx 1.48 \times 10^{-3}~\text{eV}^2$,
a point well within the reach of current experimental bounds.

For DUNE, the consequences are more severe. Using a full GLoBES
implementation with an 8-rule CPT-extended treatment that correctly
accounts for both the right-sign and wrong-sign beam contributions in
each horn polarity, we showed two qualitatively distinct failure modes
of a CPT-conserving analysis. First, for a truth with genuine CP
violation ($\delta_{\rm CP}^{\rm true} = -112^\circ$), a CPT
asymmetry $\delta\Delta m^2_{31} \lesssim -0.3 \times 10^{-3}~\text{eV}^2$
— well within current bounds — is sufficient to cause the
CPT-conserving fit to miss the true CP phase entirely, with the
inferred $\delta_{\rm CP}^{\rm meas}$ drifting by tens of degrees.
Second, and more strikingly, for a CP-conserving truth
($\delta_{\rm CP}^{\rm true} = 0^\circ$), a CPT asymmetry
$\delta\Delta m^2_{31} \gtrsim +0.76 \times 10^{-3}~\text{eV}^2$
triggers a basin switch in the $\chi^2$ landscape: the global minimum
jumps from near $\delta_{\rm CP} \approx 0^\circ$ to near
$\delta_{\rm CP} \approx \pm\pi$, producing a false detection of
near-maximal CP violation from a CP-conserving truth. In both cases
the measured $\delta_{\rm CP}$ follows the analytic $\langle\Delta
P\rangle$ isocontours on either side of the discontinuity, confirming
that the bias is a genuine feature of the degeneracy structure rather
than a numerical artifact.

The resolution lies outside the long-baseline program itself.
Atmospheric neutrino telescopes are the natural tool: their inclusive
flux contains both neutrinos and antineutrinos across a broad range of
energies and baselines, giving them direct sensitivity to
$\delta\Delta m^2_{31}$ in a single combined fit. Using existing data
from IceCube-DeepCore (8 years) and KM3NeT/ORCA-6 (433 kt-yr), we
obtain the current world-leading constraint $|\delta\Delta m^2_{31}|
\leq 0.57 \times 10^{-3}~\text{eV}^2$ at 90\% CL. With the IceCube
Upgrade (10 years from 2026) and the full KM3NeT/ORCA detector (5
years from 2031), this bound will reach $10^{-4}~\text{eV}^2$ at 90\%
CL, shrinking the allowed CPT-violating parameter space to a level
where no impostor solution can produce a $\delta_{\rm CP}^{\rm meas}$
that differs significantly from the true CP phase. The combined
IceCube+ORCA significance against the current best-fit CPT-violating
point at $\delta\Delta m^2_{31} = 5 \times 10^{-5}~\text{eV}^2$
exceeds $1\sigma$ within a decade.

The broader lesson is conceptual as much as practical. The $\Delta P$
observable accessible to any single long-baseline experiment is not a
function of $\delta_{\rm CP}$ alone: it is a function of the full
truth $(\delta_{\rm CP}, \delta\Delta m^2_{31})$, and the two
parameters are entangled on a continuous manifold. The CPT-conserving
assumption does not extract $\delta_{\rm CP}$ from that manifold; it
selects one particular point on it. Only by combining the spectral
information of long-baseline experiments with independent external
constraints on $\delta\Delta m^2_{31}$ from atmospheric telescopes can
the degeneracy be broken and the CP measurement be trusted. The
upcoming generation of experiments — DUNE, the IceCube Upgrade, and
full KM3NeT/ORCA — is precisely the combination needed to achieve
this, provided their results are analyzed jointly rather than
sequentially under the assumption that CPT is conserved.

%% file: Sections/Appendix-acc.tex
This appendix shows additional figures for the accelerator experiments analysis performed in the main text of this article to help understand the induced shift in $\dCP^\textrm{meas}$  by CPT violation and the associated impostor solution when assuming CPT invariance. 

Supp.~Fig.~1 shows the CPT-induced bias on $\delta_{\rm CP}$ at
NO$\nu$A, analogous to Fig.~4 of the main text for DUNE. The NOvA
configuration uses the stock GLoBES 4-rule setup, where each rule
is single-polarity and no rule splitting is required for CPT truth
generation. This is because the provided configuration in GLoBES~\cite{Ambats:2004js,Yang:2004NOvA,Huber_2007} includes separate rules for the right sign and wrong sign components of each flux, enabling a direct distinction on the treatment for neutrino and antineutrino fluxes in each current.  The qualitative structure — a systematic drift of the
inferred $\delta_{\rm CP}$ that tracks the $\langle\Delta P\rangle$
isocontours — is the same as for DUNE, confirming that the CP--CPT
degeneracy is not an artifact of the 8-rule DUNE implementation but
a generic feature of combined-polarity long-baseline analyses.

\begin{figure}[th]
    \centering
    \includegraphics[width=0.8\linewidth]{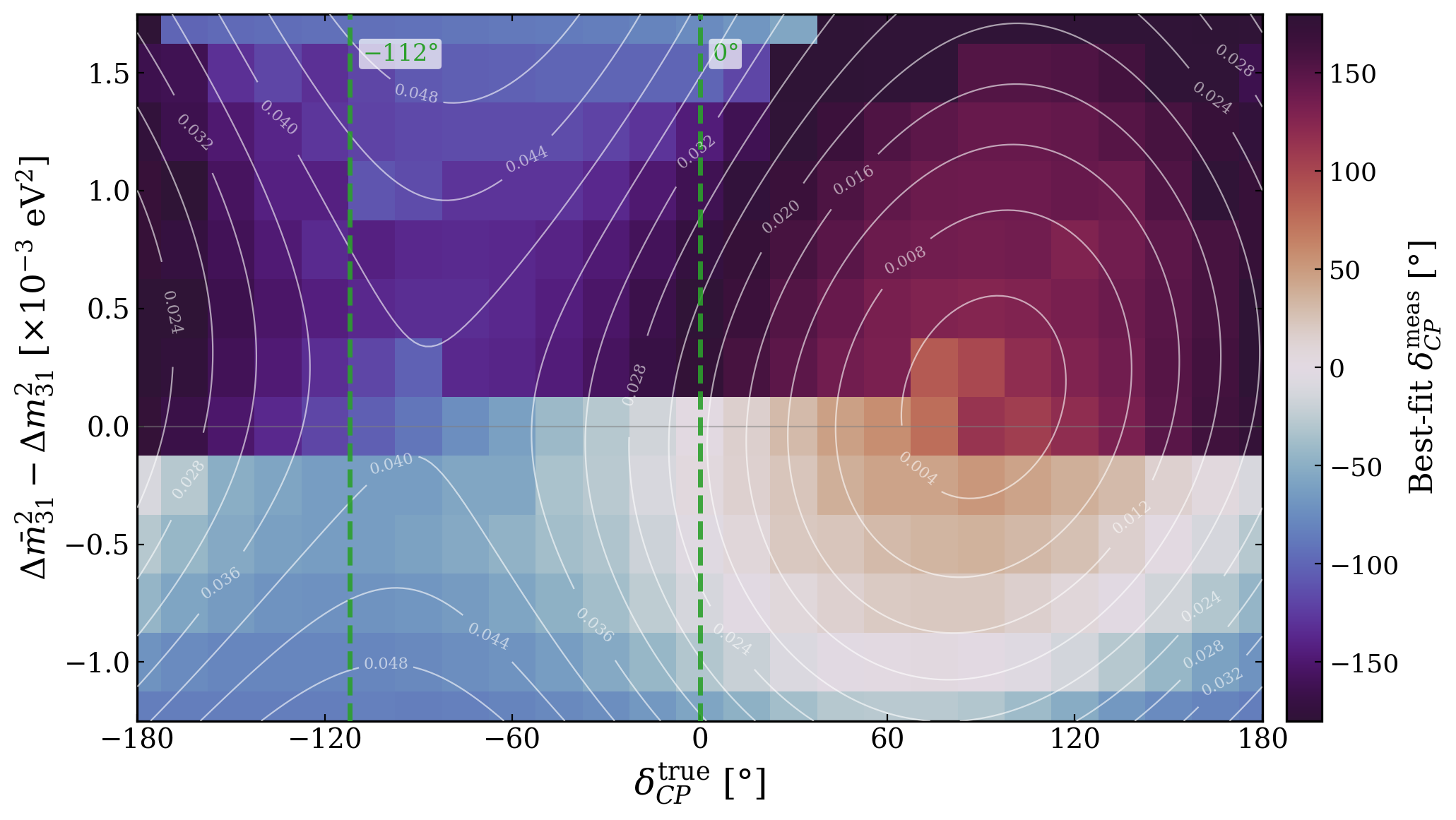}
    \caption{CPT-induced bias on $\delta_{CP}$ at NOvA (810~km, 25~kt, 3+3~yr $\nu$/$\bar\nu$). best-fit $\delta_{CP}^{\rm meas}$ is plotted as the heatmap, obtained from a CPT-conserving fit across
the truth plane
$(\delta_{CP}^{\rm true},\;\delta\Delta m^2_{31})$,
where $\delta\Delta m^2_{31}\equiv\Delta\bar{m}^2_{31}-\Delta m^2_{31}$.
White contours show the flux-weighted CP asymmetry
$\langle\Delta P\rangle
 =\langle P(\nu_\mu\!\to\!\nu_e)\rangle_{\Phi_\nu}
 -\langle P(\bar\nu_\mu\!\to\!\bar\nu_e)\rangle_{\Phi_{\bar\nu}}$
computed with the actual NuMI beam flux (separate $\nu$/$\bar\nu$ weights).
Green dashed lines mark the same two truth $\delta_{CP}$ values ($-112^\circ$ and
$0^\circ$) as in Fig.~\ref{fig:band}.
}
    \label{fig:nova_combines}
\end{figure}

Supp.~Fig.~2 provides a detailed event-level view of the CP--CPT
degeneracy at the DUNE baseline to demonstrate the behavior of the basin switch. The six panels show, for both
$\nu_e$ appearance and $\nu_\mu$ disappearance channels and both
horn polarities, the event count distributions for the CPT-violating
truth ($\delta\Delta m^2_{31} = +0.76\times10^{-3}~\text{eV}^2$,
$\delta_{\rm CP}=0^\circ$) alongside those of the two competing
CPT-conserving impostor fits. The residuals illustrate concretely
how the basin switch arises: the near-truth fit (min~1) accommodates
the $\nu_e$ appearance channel at the cost of a slight $\bar\nu_e$
mismatch, while the flipped fit (min~2) at
$\delta_{\rm CP}\approx+169^\circ$ trades a worse $\nu_e$ fit for a
better $\bar\nu_e$ fit, with the balance tipping in favor of min~2
once $|\delta\Delta|$ is large enough that the $\bar\nu_e$ mismatch
can no longer be absorbed by shifts in $\theta_{23}$ and
$\Delta m^2_{31}$. The right-most panels show the aggregated flux of both currents. This is only for the purpose of demonstrating that the two basins fit better in aggregate than the null hypothesis with CP conservation; in the actual fit the two horn currents are binned separately.

%% DUNE event distribution
\begin{figure*}
    \centering
    \includegraphics[width=\textwidth]{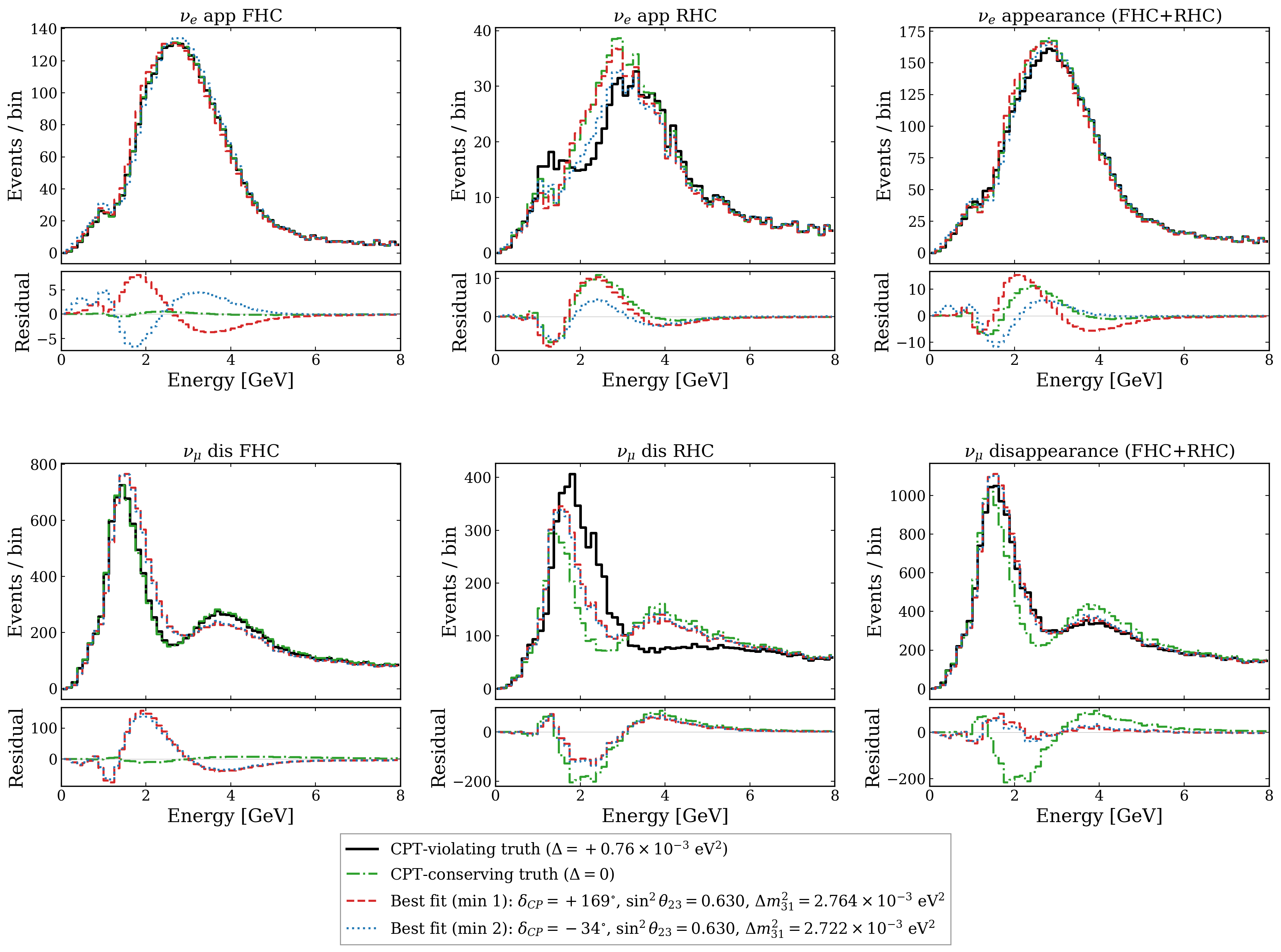}
    \caption{CP--CPT degeneracy analysis at the DUNE baseline ($L = 1284.9$~km,
$\rho = 2.848$~g/cm$^3$) for $\nu_\mu \to \nu_e$ appearance.
(a)~Decomposition of the $\nu$--$\bar\nu$ probability asymmetry
$\Delta P = P(\nu) - P(\bar\nu)$ at the truth point
($\delta_{CP} = -112^\circ$, $\Delta \equiv \Delta\bar{m}^2_{31} - \Delta m^2_{31}
= 1.0 \times 10^{-3}$~eV$^2$) into CP-violating ($\Delta P_{CP}$, blue),
CPT-violating ($\Delta P_{CPT}$, red), and matter-induced
($\Delta P_{\mathrm{matter}}$, green) contributions.
The total asymmetry (black) satisfies
$\Delta P_{\mathrm{total}} = \Delta P_{CP} + \Delta P_{CPT} + \Delta P_{\mathrm{matter}}$
by construction.
$\Delta P_{CP}$ is the vacuum asymmetry at the same mass splitting,
$\Delta P_{CPT}$ is the vacuum asymmetry difference due to the shifted
$\bar\nu$ mass splitting, and $\Delta P_{\mathrm{matter}}$ is the residual
from matter effects.
(b)~Comparison of the total asymmetry at the truth point (black solid) with
the CPT-conserving impostor ($\delta_{CP} = -84^\circ$, $\Delta = 0$;
red dashed), and their vacuum components.
(c)~Residuals between truth and impostor: $\Delta P_{\mathrm{total}} -
\Delta P'_{\mathrm{total}}$ (black) and the vacuum component difference (blue).
The shaded band marks the DUNE $\nu_e$ appearance signal window (1.3--4.7~GeV).
All probabilities computed with \texttt{nuSQuIDS} using NuFit~5.0 parameters.
}
\label{sppfig:dune_evt_dist}
\end{figure*}

%% file: Sections/Appendix-IC.tex
\begin{table*}[t]
\centering
\caption{Nuisance parameters in the combined ORCA $+$ IceCube DeepCore
CPT fit. ``Shared'' parameters enter the combined $\chi^2$ once and
multiply event weights in both experiments. ``ORCA only'' and
``IC only'' parameters are independent. The prior on every parameter
is Gaussian with the central value and width listed below.}
\label{tab:nuisance}
\begin{tabular}{lllccc}
\toprule
Group & Parameter & Application & Central $\tilde\eta_j$ & Prior $\sigma_j$ & Experiments \\
\midrule
Atmospheric flux (shared) & $\Phi_{<1\,{\rm GeV}}$        & event weight & 1.00 & 0.25 & ORCA, IC DC \\
                          & $\Phi_{>1\,{\rm GeV}}$        & event weight & 1.00 & 0.15 & ORCA, IC DC \\
                          & spectral tilt                 & event weight & 0.00 & 0.20 & ORCA, IC DC \\
                          & $\nu/\bar\nu$ ratio           & event weight & 1.00 & 0.02 & ORCA, IC DC \\
                          & $e/\mu$ ratio                 & event weight & 1.00 & 0.05 & ORCA, IC DC \\
                          & zenith (up-going)             & event weight & 0.00 & 0.20 & ORCA, IC DC \\
                          & zenith (down-going)           & event weight & 0.00 & 0.20 & ORCA, IC DC \\
\midrule
ORCA detector             & $f_{\rm all}$                 & event weight & 1.00 & 0.20 & ORCA \\
                          & $f_{\rm HPT}$                 & event weight & 1.00 & 0.20 & ORCA \\
                          & $f_{\rm Shower}$              & event weight & 1.00 & 0.20 & ORCA \\
                          & $f_{\tau{\rm CC}}$            & event weight & 1.00 & 0.20 & ORCA \\
                          & $f_{\rm NC}$                  & event weight & 1.00 & 0.20 & ORCA \\
                          & $f_{\rm HE}$                  & event weight & 1.00 & 0.50 & ORCA \\
                          & $E_{\rm shift}$               & event weight & 1.00 & 0.10 & ORCA \\
\midrule
IC DeepCore (HS)          & DOM efficiency                & bin (Eq.~\ref{eq:ic-hypersurface}) & 1.00 & 0.10 & IC DC \\
                          & hole-ice $p_0$                & bin (Eq.~\ref{eq:ic-hypersurface}) & 0.10 & 0.10 & IC DC \\
                          & hole-ice $p_1$                & bin (Eq.~\ref{eq:ic-hypersurface}) & $-0.05$ & 0.10 & IC DC \\
                          & bulk-ice absorption           & bin (Eq.~\ref{eq:ic-hypersurface}) & 1.00 & 0.10 & IC DC \\
                          & bulk-ice scattering           & bin (Eq.~\ref{eq:ic-hypersurface}) & 1.00 & 0.10 & IC DC \\
\midrule
Muon background           & $\alpha_\mu^{\rm ORCA}$       & bin (Eq.~\ref{eq:muon-add})         & 1.00 & 0.50 & ORCA \\
                          & $\alpha_\mu^{\rm IC}$         & bin (Eq.~\ref{eq:muon-add})         & 1.00 & 0.50 & IC DC \\
\bottomrule
\end{tabular}
\end{table*}

\begin{figure*}[htbp!]
\centering
\includegraphics[width=\textwidth]{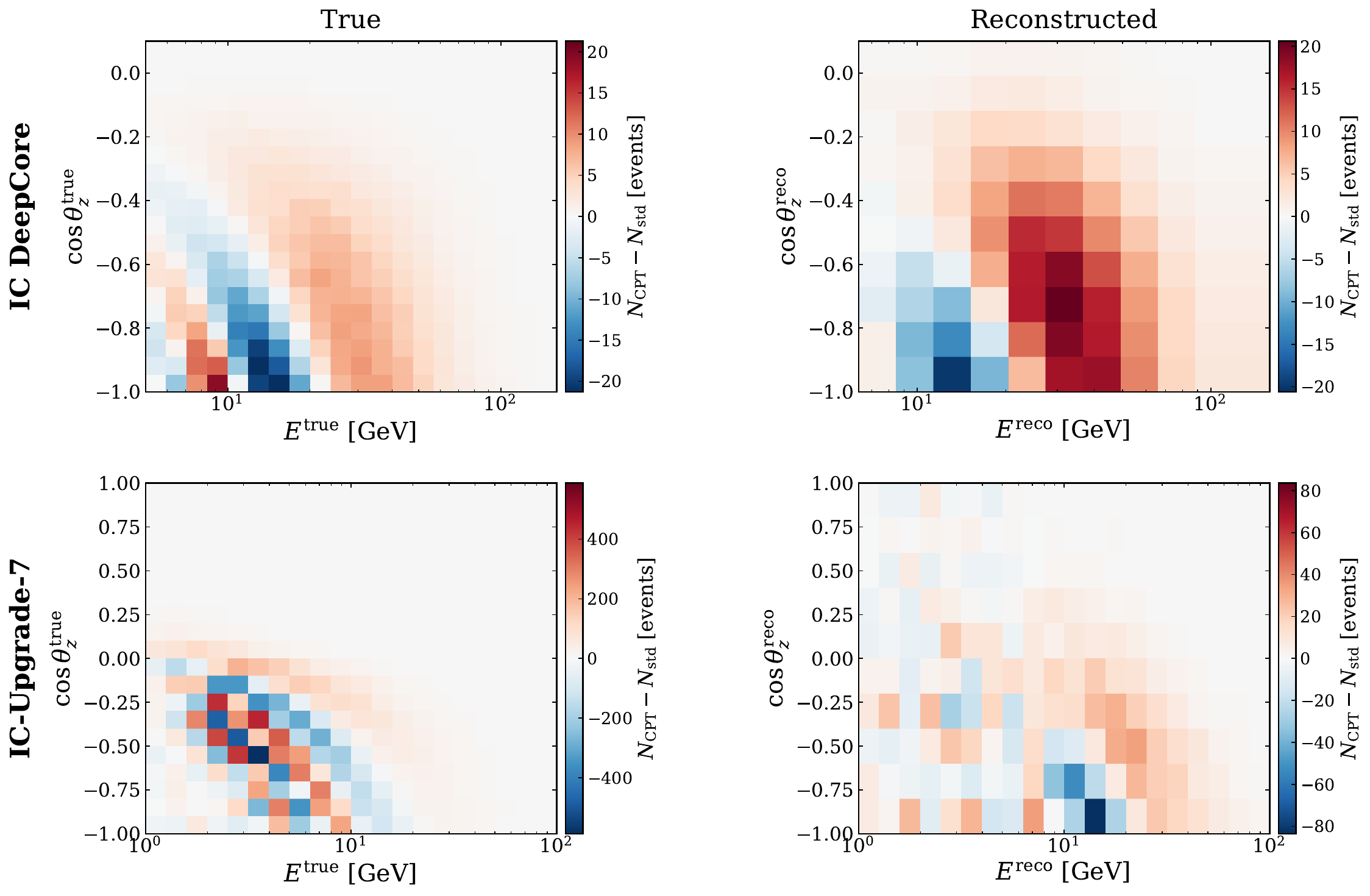}
\caption{Bin-by-bin event count difference between a CPT-violating scenario ($\Delta m^2_{31} = 2.511 \times 10^{-3}~\text{eV}^2$, $\Delta\bar{m}^2_{31} = 2.0 \times 10^{-3}~\text{eV}^2$) and the standard CPT-symmetric case for the IceCube DeepCore and IceCube-Upgrad 7-String track-like (PID~$= 1$) sample over 7.74~years of exposure. Left: true variables ($E_\text{true}$, $\cos\theta_{z}^\text{true}$). Right: reconstructed variables ($E_\text{reco}$, $\cos\theta_{z}^\text{reco}$) using respective analysis binning for the two stages of experiment. IceCube-Upgrade-7-String uses cosine zenith angle binning from upgoing to downgoing, and sees more oscillation maxima and minima patterns at lower energies, together with a finer energy and directional reconstruction.}
\label{fig:ic_true+reco}
\end{figure*}

Supp.~Table~1 shows the groups of common and experiment-specific systematic parameters used in the atmospheric analysis.
Supp.~Fig.~\ref{fig:ic_true+reco} show for IceCube DeepCore and Upgrade (7-String configuration) the true and reconstructed binned event count differences in the leading track sample between
a representative CPT-violating scenario
($\Delta m^2_{31} = 2.511\times10^{-3}~\text{eV}^2$,
$\Delta\bar{m}^2_{31} = 2.0\times10^{-3}~\text{eV}^2$) and the
standard CPT-conserving case. For IceCube DeepCore and IceCube Upgrade-7 respectively, we show the track samples' event distributions since the track sample contains more information on the oscillatory behaviors, where patterns still exist after the energy and zenith reconstruction smearing. This oscillatory pattern gives rise to the sensitivity of atmospheric telescopes to $\delta\Delta m^2_{31}$, where the granularity of the pattern after reconstruction smearing dominates the sensitivity. 

Comparing IceCube Upgrade-7 with IceCube DeepCore, the minimum neutrino energy goes down to 1 GeV, enabling the detector to see 2 more oscillation maxima, and the reconstruction improves significantly (Supp.~Figs.~5--6). Coupling this improvement to a more finely-binned event distribution histograms gives rise to the improvement to the sensitivity of CPT violation as shown in Fig.~\ref{fig:combined-fit}.

In~\ref{tab:nuisance} we also provide more information on the sources of systematic uncertainties used in this atmospheric oscillations analysis. Aside from the separate detector systematics, all cross-section systematics and flux systematics are shared between the experiments, thereby realizing a true synergetic combined fit. This prescription is similar to our previous article ~\cite{Arguelles:2022hrt}. Specifically, IceCube DeepCore employs hypersurface (HS) parametrization for detector systematics, so we also include in Eq.~\ref{eq:ic-hypersurface} a brief discussion on how these HS systematics are applied in the analysis framework.
\begin{equation}
\label{eq:ic-hypersurface}
E_i^{\rm IC}(\boldsymbol{\eta})
=
\sum_{c}
H_{c i}^{\rm IC}\!\left(\boldsymbol{\eta}_{\rm flux}\right)
\left[
a_{c i}\!\left(\Delta m^2_{31}\right)
+
\sum_{j\in{\rm HS}}
s_{c i j}\!\left(\Delta m^2_{31}\right)
\left(\eta_j-\eta_j^{\rm nom}\right)
\right],
\end{equation}
Here \(i\) labels the reconstructed analysis bin and \(c\) labels the
IceCube DeepCore hypersurface category, taken to be
\(c\in\{\mathrm{NC}+\nu_e\mathrm{CC},\,\nu_\mu\mathrm{CC},\,\nu_\tau\mathrm{CC}\}\).
The quantity \(H_{ci}^{\rm IC}(\boldsymbol{\eta}_{\rm flux})\) is the
flux-reweighted Monte Carlo expectation in bin \(i\) for category \(c\),
after applying the atmospheric-flux nuisance parameters at the event
level. The bracketed term is the hypersurface correction factor for that
category and bin. Its intercept \(a_{ci}(\Delta m^2_{31})\) gives the
nominal detector-response correction, while
\(s_{cij}(\Delta m^2_{31})\) is the linear response slope to the
IceCube detector nuisance parameter \(\eta_j\). The subtraction \(\eta_j-\eta_j^{\rm nom}\) ensures that the slope
terms vanish at the nominal hypersurface point, leaving only the
intercept correction.
The final expectation \(E_i^{\rm IC}\) is obtained by applying these
category-wise correction factors to the flux-reweighted histograms and
summing over the three categories.

\begin{figure*}[htbp!]
\centering
\includegraphics[width=0.6\textwidth]{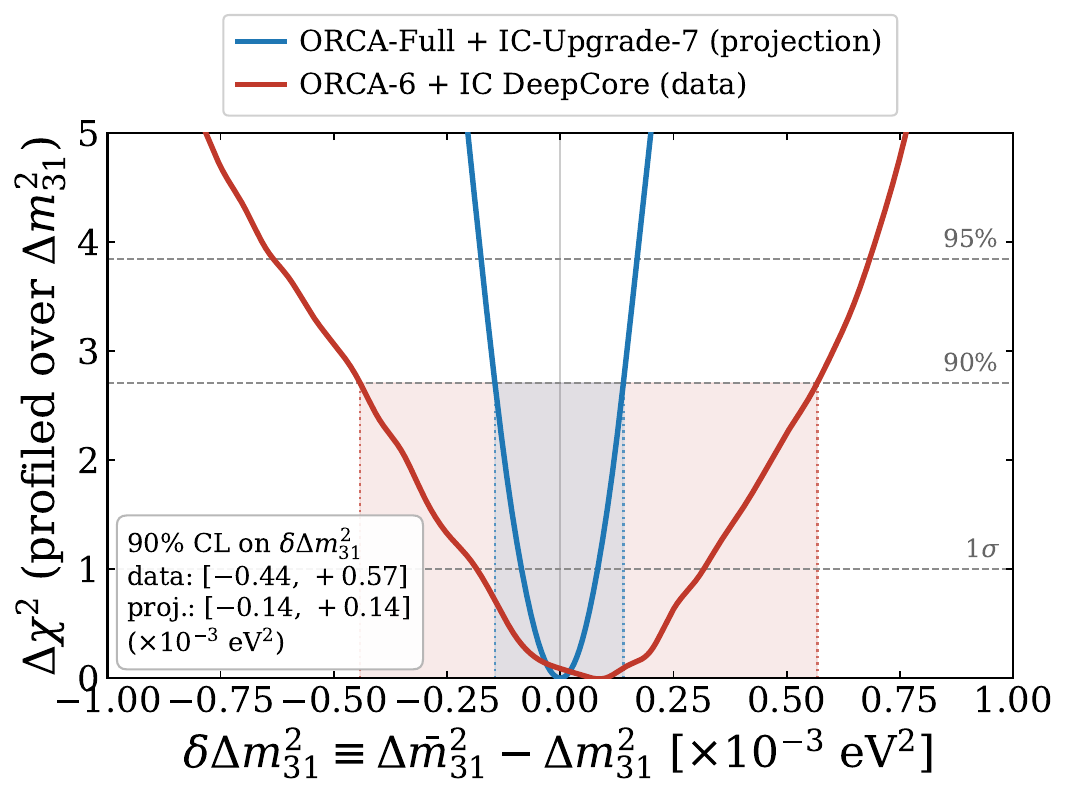}
\caption{One-dimensional profiles of the $\Delta\chi^2$ for the CPT asymmetry $\delta\Delta m^2_{31} \equiv \Delta\bar{m}^2_{31} -
  \Delta m^2_{31}$, obtained by profiling (minimizing) over $\Delta m^2_{31}$ at each value of $\delta\Delta m^2_{31}$; these are the
  one-dimensional marginalizations of the two-dimensional confidence contours of Fig.~\ref{fig:combined-fit}. \textbf{Red:} the current
  combined data fit using ORCA-6 and IceCube DeepCore, profiled over $\sin^2\theta_{23}$. \textbf{Blue:} the Asimov sensitivity projection
  for the combined ORCA-Full (5~yr) and IC-Upgrade-7 (10~yr) configuration. Horizontal dashed lines mark the $1\sigma$, 90\%, and 95\%
  confidence levels for one degree of freedom ($\Delta\chi^2 = 1.00,2.71,3.84$); the shaded bands and dotted verticals indicate the
  corresponding 90\% allowed regions. Marginalizing over $\Delta m^2_{31}$, the present data constrain $\delta\Delta m^2_{31} \in
  [-0.44,+0.57]\times10^{-3}~\mathrm{eV}^2$ at 90\% CL, fully consistent with CPT symmetry ($\delta\Delta m^2_{31}=0$), while the future
  combined configuration tightens this to $[-0.14,+0.14]\times10^{-3}~\mathrm{eV}^2$ --- an improvement of roughly a factor of $3.5$.}
\label{fig:datafit_1d}
\end{figure*}

Additionally, we report in Suppl. Fig.~\ref{fig:datafit_1d} the 1-dimensional $\chi^2$ profiles of the atmospheric CPT violation data-fit as well as the exclusion sensitivity projection.

%% file: main.bib
@article{10.1088/1361-6471/ae7818,
	author={Fernandez-Menendez, Pablo},
	title={CHIC: Caley-Hamilton, invariants and constants for neutrino oscillation probabilities and gradients},
	journal={Journal of Physics G: Nuclear and Particle Physics},
	url={http://iopscience.iop.org/article/10.1088/1361-6471/ae7818},
	year={2026}
}

@software{chic,
  author = "Fern{\'a}ndez-Men{\'e}ndez, Pablo",
  doi = {10.5281/zenodo.1234},
  month = {5},
  title = {{CHIC Neutrino Oscillations and Derivatives}},
  url = {https://github.com/pabloferm/CHIC/releases/tag/v2.0.0},
  version = {2.0.0},
  year = {2026}
}

@article{Barenboim_2018,
   title={Neutrinos, DUNE and the world best bound on CPT invariance},
   volume={780},
   ISSN={0370-2693},
   url={http://dx.doi.org/10.1016/j.physletb.2018.03.060},
   DOI={10.1016/j.physletb.2018.03.060},
   journal={Physics Letters B},
   publisher={Elsevier BV},
   author={Barenboim, G. and Ternes, C.A. and Tórtola, M.},
   year={2018},
   month=may, pages={631–637} }

@article{Huber_2005,
   title={Simulation of long-baseline neutrino oscillation experiments with GLoBES},
   volume={167},
   ISSN={0010-4655},
   url={http://dx.doi.org/10.1016/j.cpc.2005.01.003},
   DOI={10.1016/j.cpc.2005.01.003},
   number={3},
   journal={Computer Physics Communications},
   publisher={Elsevier BV},
   author={Huber, P. and Lindner, M. and Winter, W.},
   year={2005},
   month=may, pages={195–202} }

@article{Huber_2007,
   title={New features in the simulation of neutrino oscillation experiments with GLoBES 3.0},
   volume={177},
   ISSN={0010-4655},
   url={http://dx.doi.org/10.1016/j.cpc.2007.05.004},
   DOI={10.1016/j.cpc.2007.05.004},
   number={5},
   journal={Computer Physics Communications},
   publisher={Elsevier BV},
   author={Huber, Patrick and Kopp, Joachim and Lindner, Manfred and Rolinec, Mark and Winter, Walter},
   year={2007},
   month=sep, pages={432–438} }

@article{Nunokawa:2007qh,
    author = "Nunokawa, Hiroshi and Parke, Stephen J. and Valle, Jose W. F.",
    title = "{CP Violation and Neutrino Oscillations}",
    eprint = "0710.0554",
    archivePrefix = "arXiv",
    primaryClass = "hep-ph",
    reportNumber = "FERMILAB-PUB-07-521-T, IFIC-07-58",
    doi = "10.1016/j.ppnp.2007.10.001",
    journal = "Prog. Part. Nucl. Phys.",
    volume = "60",
    pages = "338--402",
    year = "2008"
}

@article{Olavarrieta:2024eaq,
    author = {Olavarrieta, Santiago Giner and Jin, Miaochen and Arg\"uelles, Carlos A. and Fern\'andez, Pablo and Mart\'\i{}nez-Soler, Ivan},
    title = "{Boosting neutrino mass ordering sensitivity with inelasticity for atmospheric neutrino oscillation measurement}",
    eprint = "2402.13308",
    archivePrefix = "arXiv",
    primaryClass = "hep-ph",
    reportNumber = "IPPP/24/05",
    doi = "10.1103/PhysRevD.110.L051101",
    journal = "Phys. Rev. D",
    volume = "110",
    number = "5",
    pages = "L051101",
    year = "2024"
}

@article{Arguelles:2022hrt,
    author = {Arg\"uelles, C. A. and Fern\'andez, P. and Mart\'\i{}nez-Soler, I. and Jin, M.},
    title = "{Measuring Oscillations with a Million Atmospheric Neutrinos}",
    eprint = "2211.02666",
    archivePrefix = "arXiv",
    primaryClass = "hep-ph",
    doi = "10.1103/PhysRevX.13.041055",
    journal = "Phys. Rev. X",
    volume = "13",
    number = "4",
    pages = "041055",
    year = "2023"
}

@article{Abbasi_2023,
   title={Measurement of atmospheric neutrino mixing with improved IceCube DeepCore calibration and data processing},
   volume={108},
   ISSN={2470-0029},
   url={http://dx.doi.org/10.1103/PhysRevD.108.012014},
   DOI={10.1103/physrevd.108.012014},
   number={1},
   journal={Physical Review D},
   publisher={American Physical Society (APS)},
   author={Abbasi, R. and Ackermann, M. and Adams, J. and Agarwalla, S. K. and others},
   year={2023},
   month=jul }

@article{km3netcollaboration2024measurementneutrinooscillationparameters,
    author = "Aiello, S. and others",
    collaboration = "KM3NeT",
    title = "{Measurement of neutrino oscillation parameters with the first six detection units of KM3NeT/ORCA}",
    eprint = "2408.07015",
    archivePrefix = "arXiv",
    primaryClass = "hep-ex",
    doi = "10.1007/JHEP10(2024)206",
    journal = "JHEP",
    volume = "10",
    pages = "206",
    year = "2024",
    note = "[Addendum: JHEP 10, 041 (2025)]"
}

@misc{dunecollaboration2016longbaselineneutrinofacilitylbnf,
      title={Long-Baseline Neutrino Facility (LBNF) and Deep Underground Neutrino Experiment (DUNE) Conceptual Design Report Volume 2: The Physics Program for DUNE at LBNF}, 
      author={DUNE Collaboration},
      year={2016},
      eprint={1512.06148},
      archivePrefix={arXiv},
      primaryClass={physics.ins-det},
      url={https://arxiv.org/abs/1512.06148}, 
}

@article{Barenboim:2001ac,
  author        = {Barenboim, G. and Borissov, L. and Lykken, J. and Smirnov, A. Yu.},
  title         = {Neutrinos as the messengers of {CPT} violation},
  journal       = {JHEP},
  volume        = {10},
  pages         = {001},
  year          = {2002},
  doi           = {10.1088/1126-6708/2002/10/001},
  eprint        = {hep-ph/0108199},
  archivePrefix = {arXiv}
}

@article{Barenboim:2020tzf,
  author        = {Barenboim, G. and Ternes, C. A. and T{\'o}rtola, M.},
  title         = {{CPT} and {CP}, an entangled couple},
  journal       = {JHEP},
  volume        = {07},
  pages         = {155},
  year          = {2020},
  doi           = {10.1007/JHEP07(2020)155},
  eprint        = {2005.05975},
  archivePrefix = {arXiv},
  primaryClass  = {hep-ph}
}

@article{Fukuda:1998mi,
  author        = {Fukuda, Y. and others},
  collaboration = {Super-Kamiokande},
  title         = {Evidence for an anomalous component of cosmic-ray muons},
  journal       = {Phys. Rev. Lett.},
  volume        = {81},
  pages         = {1562--1567},
  year          = {1998},
  doi           = {10.1103/PhysRevLett.81.1562},
  eprint        = {hep-ex/9807003},
  archivePrefix = {arXiv}
}

@article{Ahmad:2002jz,
  author        = {Ahmad, Q. R. and others},
  collaboration = {SNO},
  title         = {Direct evidence for neutrino flavor transformation from neutral-current
                   interactions in the Sudbury Neutrino Observatory},
  journal       = {Phys. Rev. Lett.},
  volume        = {89},
  pages         = {011301},
  year          = {2002},
  doi           = {10.1103/PhysRevLett.89.011301},
  eprint        = {nucl-ex/0204008},
  archivePrefix = {arXiv}
}

@misc{Ambats:2004js,
      author        = "Ambats, I. and others",
      collaboration = "NOvA",
      title         = "{NOvA proposal to build a 30-kiloton off-axis detector to
                       study neutrino oscillations in the Fermilab NuMI beamline}",
       eprint        = "hep-ex/0503053",
      archivePrefix = "arXiv",                                                  
      reportNumber  = "FERMILAB-PROPOSAL-0929",
      year          = "2004"                                                    
  }

@techreport{Yang:2004NOvA,                                                    
  author        = "Yang, T. and Wojcicki, S.",
  collaboration = "NOvA",                                                   
  title         = "{Study of physics sensitivity of $\nu_\mu$ disappearance
                   in a totally active version of NOvA detector}",          
  institution   = "NOvA Collaboration",
  number        = "Off-Axis-Note-SIM-30",                                   
  year          = "2004"                                                    
}

@article{T2K:2025wet,
    author = "Abubakar, S. and others",
    collaboration = "T2K, NOvA",
    title = "{Joint neutrino oscillation analysis from the T2K and NOvA experiments}",
    eprint = "2510.19888",
    archivePrefix = "arXiv",
    primaryClass = "hep-ex",
    reportNumber = "FERMILAB-PUB-25-0132-PPD",
    doi = "10.1038/s41586-025-09599-3",
    journal = "Nature",
    volume = "646",
    number = "8086",
    pages = "818--824",
    year = "2025"
}

@article{DUNE:2020jqi,
  author        = {Abi, B. and others},
  collaboration = {DUNE},
  title         = {Long-baseline neutrino oscillation physics potential of the {DUNE} experiment},
  journal       = {Eur. Phys. J. C},
  volume        = {80},
  pages         = {978},
  year          = {2020},
  doi           = {10.1140/epjc/s10052-020-08456-3},
  eprint        = {2006.16043},
  archivePrefix = {arXiv},
  primaryClass  = {hep-ex}
}

@article{Luders:1954zz,
  author  = {L{\"u}ders, G.},
  title   = {On the equivalence of invariance under time reversal and under
             particle-antiparticle conjugation for relativistic field theories},
  journal = {Kong. Dan. Vid. Sel. Mat. Fys. Med.},
  volume  = {28N5},
  pages   = {1--17},
  year    = {1954}
}

@inbook{Pauli:1955,
  author    = {Pauli, W.},
  title     = {Exclusion principle, {L}orentz group and reflection of space-time and charge},
  booktitle = {Niels Bohr and the Development of Physics},
  editor    = {Pauli, W.},
  publisher = {McGraw-Hill},
  address   = {New York},
  year      = {1955},
  pages     = {30--51}
}

@article{Bell:1955,
  author  = {Bell, J. S.},
  title   = {Time reversal in field theory},
  journal = {Proc. Roy. Soc. Lond. A},
  volume  = {231},
  pages   = {479--495},
  year    = {1955},
  doi     = {10.1098/rspa.1955.0189}
}

@article{Jost:1957zz,
  author  = {Jost, R.},
  title   = {Eine Bemerkung zum {CTP} Theorem},
  journal = {Helv. Phys. Acta},
  volume  = {30},
  pages   = {409--416},
  year    = {1957}
}

@article{Kostelecky:1988zi,
  author  = {Kosteleck{\'y}, V. A. and Samuel, S.},
  title   = {Spontaneous breaking of {L}orentz symmetry in string theory},
  journal = {Phys. Rev. D},
  volume  = {39},
  pages   = {683},
  year    = {1989},
  doi     = {10.1103/PhysRevD.39.683}
}

@article{Kostelecky:1991ak,
  author  = {Kosteleck{\'y}, V. A. and Samuel, S.},
  title   = {Gravitational phenomenology in higher-dimensional theories and strings},
  journal = {Phys. Rev. D},
  volume  = {40},
  pages   = {1886},
  year    = {1989},
  doi     = {10.1103/PhysRevD.40.1886}
}

@techreport{KM3NeT_IN2P3_status_2026,
    author      = {{KM3NeT Collaboration}},
    title       = {Report on the Status of {KM3NeT} for the {IN2P3} Scientific Council},
    institution = {APC, CPPM, IPHC, LPC, Subatech, LUPM},
    type        = {Status Report},
    year        = {2026},
    month       = mar,
    note        = {Full KM3NeT/ORCA detector completion expected by 2030},
    url         =
  {https://atrium.in2p3.fr/nuxeo/nxfile/default/57521610-caf6-43ec-ad07-ce61b4f90f06/blobholder:0/CSIN2P3_KM3NeT_Feb2026.pdf}
  }

@article{Colladay:1996iz,
  author        = {Colladay, D. and Kosteleck{\'y}, V. A.},
  title         = {{CPT} violation and the standard model},
  journal       = {Phys. Rev. D},
  volume        = {55},
  pages         = {6760--6774},
  year          = {1997},
  doi           = {10.1103/PhysRevD.55.6760},
  eprint        = {hep-ph/9703464},
  archivePrefix = {arXiv}
}

@article{Colladay:1998fh,
  author        = {Colladay, D. and Kosteleck{\'y}, V. A.},
  title         = {Lorentz-violating extension of the standard model},
  journal       = {Phys. Rev. D},
  volume        = {58},
  pages         = {116002},
  year          = {1998},
  doi           = {10.1103/PhysRevD.58.116002},
  eprint        = {hep-ph/9809521},
  archivePrefix = {arXiv}
}

@article{Kostelecky:2003fs,
  author        = {Kosteleck{\'y}, V. A.},
  title         = {Gravity, {L}orentz violation, and the standard model},
  journal       = {Phys. Rev. D},
  volume        = {69},
  pages         = {105009},
  year          = {2004},
  doi           = {10.1103/PhysRevD.69.105009},
  eprint        = {hep-th/0312310},
  archivePrefix = {arXiv}
}

@article{Amelino-Camelia:1997ieq,
  author        = {Amelino-Camelia, G. and Ellis, J. R. and Mavromatos, N. E.
                   and Nanopoulos, D. V. and Sarkar, S.},
  title         = {Tests of quantum gravity from observations of gamma-ray bursts},
  journal       = {Nature},
  volume        = {393},
  pages         = {763--765},
  year          = {1998},
  doi           = {10.1038/31647},
  eprint        = {astro-ph/9712103},
  archivePrefix = {arXiv}
}

@article{Mavromatos:2004sz,
  author        = {Mavromatos, N. E.},
  title         = {{CPT} violation and decoherence in quantum gravity},
  journal       = {Lect. Notes Phys.},
  volume        = {669},
  pages         = {245--320},
  year          = {2005},
  doi           = {10.1007/11377306_8},
  eprint        = {gr-qc/0407005},
  archivePrefix = {arXiv}
}

@article{Kostelecky:2003cr,
  author        = {Kosteleck{\'y}, V. A. and Mewes, M.},
  title         = {Lorentz and {CPT} violation in neutrinos},
  journal       = {Phys. Rev. D},
  volume        = {69},
  pages         = {016005},
  year          = {2004},
  doi           = {10.1103/PhysRevD.69.016005},
  eprint        = {hep-ph/0309025},
  archivePrefix = {arXiv}
}

@article{IceCube:2010fyu,
  author        = {Abbasi, R. and others},
  collaboration = {IceCube},
  title         = {Search for a {L}orentz-violating sidereal signal with atmospheric neutrinos
                   in {IceCube}},
  journal       = {Phys. Rev. D},
  volume        = {82},
  pages         = {112003},
  year          = {2010},
  doi           = {10.1103/PhysRevD.82.112003},
  eprint        = {1010.4096},
  archivePrefix = {arXiv},
  primaryClass  = {astro-ph.HE}
}

@misc{conway2011incorporatingnuisanceparameterslikelihoods,
      title={Incorporating Nuisance Parameters in Likelihoods for Multisource Spectra}, 
      author={J. S. Conway},
      year={2011},
      eprint={1103.0354},
      archivePrefix={arXiv},
      primaryClass={physics.data-an},
      url={https://arxiv.org/abs/1103.0354}, 
}

@article{Barlow:1993jj,
    author         = "Barlow, Roger J. and Beeston, Christine",
    title          = "{Fitting using finite Monte Carlo samples}",
    journal        = "Comput. Phys. Commun.",
    volume         = "77",
    year           = "1993",
    pages          = "219-228",
    doi            = "10.1016/0010-4655(93)90005-W",
    reportNumber   = "MAN-HEP-93-1",
    SLACcitation   = "%%CITATION = CPHCB,77,219;%%"
}

@misc{ICupMC,
  author        = {{IceCube Collaboration}},
  title         = {{IceCube Upgrade Monte Carlo}},
  doi           = {10.21234/qfz1-yh02},
  year          = {2024}
}

@misc{ORCA-data,
  author        = {{KM3NeT Collaboration}},
  title         = {{ORCA6 Data Release}},
  url           = {https://zenodo.org/records/15715220},
  year          = {2025}
}

@misc{Abe2018Hyper-KamiokandeReport,
  author        = {Abe, K. and others},
  collaboration = {Hyper-Kamiokande Proto-Collaboration},
  title         = {{Hyper-Kamiokande Design Report}},
  eprint        = {1805.04163},
  archivePrefix = {arXiv},
  primaryClass  = {physics.ins-det},
  year          = {2018},
  month         = may
}

@article{Arguelles2023MeasuringNeutrinos,
    title = {{Measuring Oscillations with a Million Atmospheric Neutrinos}},
    year = {2023},
    journal = {Phys. Rev. X},
    author = {Arg{\"{u}}elles, C A and Fern{\'{a}}ndez, P and Mart{\textbackslash}'{\textbackslash}inez-Soler, I and Jin, M},
    number = {4},
    pages = {41055},
    volume = {13},
    doi = {10.1103/PhysRevX.13.041055},
    arxivId = {2211.02666}
}

@article{Cervera:2000kp,
    author = "Cervera, A. and Donini, A. and Gavela, M. B. and Gomez Cadenas, J. J. and Hernandez, P. and Mena, Olga and Rigolin, S.",
    title = "{Golden measurements at a neutrino factory}",
    eprint = "hep-ph/0002108",
    archivePrefix = "arXiv",
    reportNumber = "CERN-TH-2000-040, FTUAM-00-03, IFT-UAM-CSIC-00-04, FTUV-00-12, IFIC-00-13",
    doi = "10.1016/S0550-3213(00)00221-2",
    journal = "Nucl. Phys. B",
    volume = "579",
    pages = "17--55",
    year = "2000",
    note = "[Erratum: Nucl.Phys.B 593, 731--732 (2001)]"
}

@misc{Beacom:2026mys,
    author = "Beacom, John F. and Bell, Nicole F. and Dolan, Matthew J. and Meighen-Berger, Stephan A. and Yim, Ho Man",
    title = "{Towards Measuring the CP-Violating Phase with Atmospheric Neutrinos}",
    eprint = "2605.16721",
    archivePrefix = "arXiv",
    primaryClass = "hep-ph",
    month = "5",
    year = "2026"
}

@article{Denton:2023qmd,
    author = "Denton, Peter B.",
    title = "{Probing CP Violation with Neutrino Disappearance Alone}",
    eprint = "2309.03262",
    archivePrefix = "arXiv",
    primaryClass = "hep-ph",
    doi = "10.1103/PhysRevLett.133.031801",
    journal = "Phys. Rev. Lett.",
    volume = "133",
    number = "3",
    pages = "031801",
    year = "2024"
}

@article{Razzaque:2014vba,
    author = "Razzaque, Soebur and Smirnov, A. Yu.",
    title = "{Super-PINGU for measurement of the leptonic CP-phase with atmospheric neutrinos}",
    eprint = "1406.1407",
    archivePrefix = "arXiv",
    primaryClass = "hep-ph",
    doi = "10.1007/JHEP05(2015)139",
    journal = "JHEP",
    volume = "05",
    pages = "139",
    year = "2015"
}

@article{Kelly:2019itm,
    author = "Kelly, Kevin James and Machado, Pedro AN and Martinez Soler, Ivan and Parke, Stephen J and Perez Gonzalez, Yuber F",
    title = "{Sub-GeV Atmospheric Neutrinos and CP-Violation in DUNE}",
    eprint = "1904.02751",
    archivePrefix = "arXiv",
    primaryClass = "hep-ph",
    reportNumber = "FERMILAB-PUB-19-136-T, NUHEP-TH/19-03",
    doi = "10.1103/PhysRevLett.123.081801",
    journal = "Phys. Rev. Lett.",
    volume = "123",
    number = "8",
    pages = "081801",
    year = "2019"
}

@article{Ternes:2019sak,
    author = "Ternes, Christoph A. and Gariazzo, Stefano and Hajjar, Rasmi and Mena, Olga and Sorel, Michel and T{\'o}rtola, Mariam",
    title = "{Neutrino mass ordering at DUNE: An extra $\nu$ bonus}",
    eprint = "1905.03589",
    archivePrefix = "arXiv",
    primaryClass = "hep-ph",
    doi = "10.1103/PhysRevD.100.093004",
    journal = "Phys. Rev. D",
    volume = "100",
    number = "9",
    pages = "093004",
    year = "2019"
}

@article{Martinez-Soler:2019nhb,
    author = "Martinez-Soler, Ivan and Minakata, Hisakazu",
    title = "{Perturbing Neutrino Oscillations Around the Solar Resonance}",
    eprint = "1904.07853",
    archivePrefix = "arXiv",
    primaryClass = "hep-ph",
    reportNumber = "FERMILAB-PUB-19-149-T, NUHEP-TH/19-05",
    doi = "10.1093/ptep/ptz067",
    journal = "PTEP",
    volume = "2019",
    number = "7",
    pages = "073B07",
    year = "2019"
}

@misc{DUNE:2020ypp,
    author = "Abi, Babak and others",
    collaboration = "DUNE",
    title = "{Deep Underground Neutrino Experiment (DUNE), Far Detector Technical Design Report, Volume II: DUNE Physics}",
    eprint = "2002.03005",
    archivePrefix = "arXiv",
    primaryClass = "hep-ex",
    reportNumber = "FERMILAB-PUB-20-025-ND, FERMILAB-DESIGN-2020-02",
    doi = "10.2172/1599307",
    month = "2",
    year = "2020"
}
